\documentclass[preprint2]{proto}
\usepackage{natbib}
\bibpunct{(}{)}{;}{a}{,}{,}
\usepackage{times}
\usepackage{graphicx}
\usepackage{textcomp}
\usepackage{array}

\begin{document}

\title{\textbf{\LARGE Deuterium Fractionation: the Ariadne's Thread from
    the Pre-collapse Phase to Meteorites and Comets today}}

\author {\textbf{\large Cecilia Ceccarelli}}
\affil{\small\em Universit\'e J.Fourier de Grenoble}
\author {\textbf{\large Paola Caselli}}
\affil{\small\em University of Leeds}
\author {\textbf{\large Dominique Bockel\'ee-Morvan}}
\affil{\small\em  Observatoire de Paris}
\author {\textbf{\large Olivier Mousis}}
\affil{\small\em Universit\'e Franche-Comt\'e and Universit\'e de Toulouse}
\author {\textbf{\large Sandra Pizzarello}}
\affil{\small\em Arizona State University}
\author {\textbf{\large Francois Robert}}
\affil{\small\em Mus\'eum National d'Histoire Naturelle de Paris}
\author {\textbf{\large Dmitry Semenov}}
\affil{\small\em Max Planck Institute for Astronomy, Heidelberg}

\begin{abstract}
\baselineskip = 11pt
\leftskip = 0.65in 
\rightskip = 0.65in
\parindent=1pc {\small The Solar System formed about 4.6 billion years
  ago from a condensation of matter inside a molecular cloud. Trying
  to reconstruct what happened is the goal of this chapter. For that,
  we put together our understanding of Galactic objects that will
  eventually form new suns and planetary systems, with our knowledge
  on comets, meteorites and small bodies of the Solar System
  today. Our specific tool is the molecular deuteration, namely
  the amount of deuterium with respect to hydrogen in molecules. This
  is the Ariadne's thread that helps us to find the way out from a
  labyrinth of possible histories of our Solar System. The chapter
  reviews the observations and theories of the deuterium fractionation
  in pre-stellar cores, protostars, protoplanetary disks, comets,
  interplanetary dust particles and meteorites and links them together
  trying to build up a coherent picture of the history of the Solar
  System formation. We emphasise the interdisciplinary nature of the
  chapter, which gathers together researchers from different
  communities with the common goal of understanding the Solar System
  history. 
  \\~\\~\\~}

\end{abstract}

\section{\textbf{INTRODUCTION}}\label{sec:introduction}

The ancient Greek legend reads that Theseus volunteered to enter in
the Minotaur's labyrinth to kill the monster and liberate Athens from
periodically providing young women and men in sacrifice. The task was
almost impossible to achieve because killing the Minotaur was not even
half of the problem: getting out of the labyrinth was even more
difficult. But Ariadne, the guardian of the labyrinth and daughter of
the king of Crete, provided Theseus with a ball of thread, so that he
could unroll it going inside and follow it back to get out of the
Minotaur's labyrinth. Which he did.

Our labyrinth here is the history of the formation of the Solar
System. We are deep inside the labyrinth, with the Earth and the
planets formed, but we don't know how exactly this happened. There are
several paths that go into different directions, but what is the one
that will bring us out of this labyrinth, the path Nature followed to
form the Earth and the other Solar System planets and bodies?

Our story reads that once upon a time, it existed an interstellar
cloud of gas and dust.
Then, about 4.6 billions years ago, one cloud fragment became the Solar
System. What happened to that primordial condensation? When, why and
how did it happen?  Answering these questions involves putting
together all of the information we have on the present day Solar
System bodies and micro particles. But this is not enough, and
comparing that information with our understanding of the formation
process of Solar-type stars in our Galaxy turns out to be
indispensable too.

Our Ariadne's thread for this chapter is the deuterium fractionation,
namely the process that enriches the amount of deuterium with respect
to hydrogen in molecules. Although deuterium atoms are only
$\sim1.6\times 10^{-5}$ (Tab. \ref{tab:definitions}) times as abundant as the
hydrogen atoms in the Universe, its relative abundance in molecules,
larger than the elemental D/H abundance in very specific situations,
provides a remarkable and almost unique diagnostic tool.
Analysing the deuterium fractionation in different galactic objects
which will eventually form new suns, and in comets, meteorites and
small bodies of the Solar System is like having in our hands a box of
old photos with the imprint of memories, from the very first steps of
the Solar System formation. The goal of this chapter is
trying to understand the message that these photos bring, using our
knowledge of the different objects and, in particular, the Ariadne's
thread of the deuterium fractionation to link them together in a
sequence that is the one that followed the Solar System formation.

The chapter is organised as follows. In \S \ref{sec:set-stage}, we
review the mechanisms of the deuterium fractionation in the different
environments and set the bases for understanding the language of the
different communities involved in the study of the Solar System
formation. We then briefly review the major steps of the formation
process in \S \ref{sec:brief-hist}. The following sections, from \S
\ref{sec:the-pre-stell} to \S \ref{sec:solar-nebula}, will review in
detail observations and theories of deuterium fractionation in the
different objects: pre-stellar cores, protostars, protoplanetary disks,
comets and meteorites. In \S \ref{sec:summaryD}, we will try to follow
back the thread, unrolled in the precedent sections, to understand
what happened to the Solar System, including the formation of the
terrestrial oceans. We will conclude with \S \ref{sec:conclusions}.


\section{\textbf{THE LANGUAGE OF DEUTERIUM FRACTIONATION}}\label{sec:set-stage}

\subsection{\textbf{Chemical processes of deuterium
    fractionation}}\label{sec:the-chem-proc}

\begin{figure}[bt]
  \begin{center}
  \includegraphics[width=9cm]{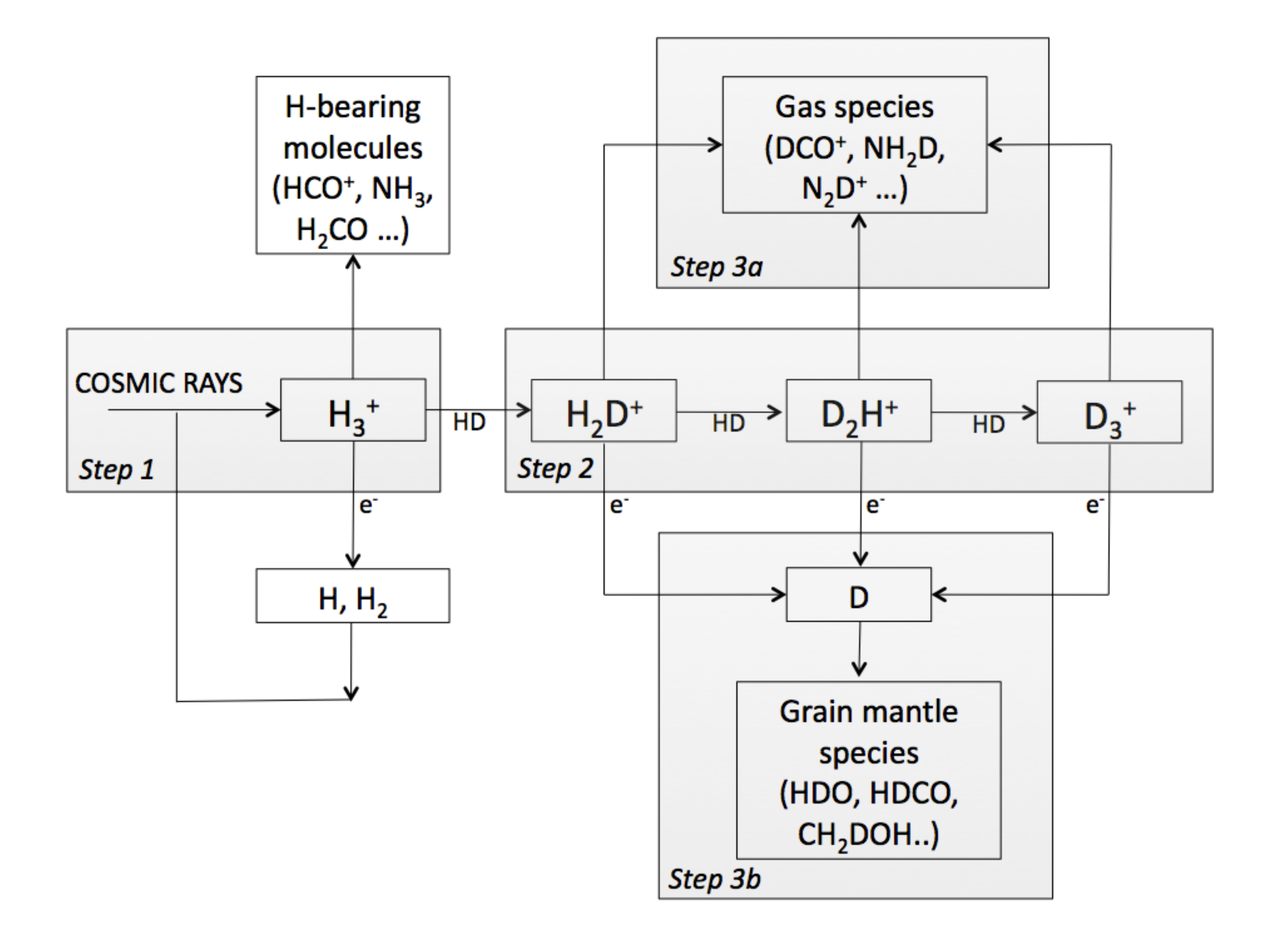}
  \caption{\small  In cold ($\leq 30$K) gas, deuterium fractionation occurs
    through three basic steps: 1) formation of H$_3^+$ ions from the
    interaction of cosmic rays with H and H$_2$; 2) formation of
    H$_2$D$^+$ (HD$_2^+$ and D$_3^+$) from the reaction of H$_3^+$
    (H$_2$D$^+$ and HD$_2^+$) with HD; 3) formation of other D-bearing
    molecules from reactions with H$_2$D$^+$ (HD$_2^+$ and
    D$_3^+$) in the gas phase (Step 3a) and on the grain mantles (Step
    3b).} \label{fig:D-chemistry-scheme}
  \end{center}
\end{figure}
Deuterium is formed at the birth of the Universe with an abundance D/H
estimated to be $A_D = 1.6 \times 10^{-5}$
(Tab. \ref{tab:definitions}) and destroyed in the interiors of the
stars. Therefore, its abundance may vary from place to place: for
example, it is lower in regions close to the Galactic Center,
where the star formation is high, than in the Solar System
neighborhood \citep{2000Natur.405.1025L}. If there were no deuterium
fractionation, a species with one H-atom, like for example HCN, would
have a relative abundance of D-atom over H-atom bearing molecules
equal to $A_D$, namely DCN/HCN$= 1.6 \times
10^{-5}$. As another important example, water would have
HDO/H$_2$O=$2\times A_D = 3.2 \times 10^{-5}$. Similarly, a species
with two hydrogens will have a relative abundance of molecules with
two D-atoms proportional to $A_D^2$ (e.g., D$_2$O/H$_2$O$= 2.6 \times
10^{-10}$) and so on. In practice, if there were no deuterium
fractionation, the abundance of D-bearing molecules would be
ridiculously low.

But in space things are special enough to make the conditions
propitious for deuterium fractionation (or molecular deuteration or
deuterium enrichment) to occur. This can be summarised in three basic
steps,
shown in Fig. \ref{fig:D-chemistry-scheme}:\\
{\it 1) formation of H$_3^+$:} in cold ($\sim 10$ K) molecular gas,
the fastest reactions are those involving ions, as neutral-neutral
reactions have activation barriers and are generally slower. The first
formed molecular ion is H$_3^+$, a product of the cosmic rays
ionisation of H$_2$ and H.
\\
{\it 2) Formation of H$_2$D$^+$, D$_2$H$^+$ and D$_3^+$:} in cold
molecular gas, H$_3^+$ reacts with HD, the major reservoir of D-atoms,
and once every three times the D-atom is transfered from HD to
H$_2$D$^+$. The inverse reaction H$_2$ + H$_2$D$^+$ which would form
HD has a (small) activation barrier so that at low temperatures
H$_2$D$^+$/H$_3^+$ becomes larger than $A_D$. Similarly, D$_2$H$^+$
and D$_3^+$ are formed by reactions with HD.
\\
{\it 3) Formation of other D-bearing molecules:} H$_2$D$^+$,
D$_2$H$^+$ and D$_3^+$ react with other molecules and atoms
transferring the D-atoms to all the other species. This can happen
directly in the gas phase ({\it Step 3a} in
Fig. \ref{fig:D-chemistry-scheme}) or on the grain mantles ({\it Step
  3b}) via the D atoms created by the H$_2$D$^+$, D$_2$H$^+$ and
D$_3^+$ dissociative recombination with electrons. In both cases, the
deuterium fractionation depends on the H$_2$D$^+$/H$_3^+$,
D$_2$H$^+$/H$_3^+$ and D$_3^+$/H$_3^+$ abundance ratios.
\begin{figure*}[bt]
   \begin{center}
 \includegraphics[width=12cm]{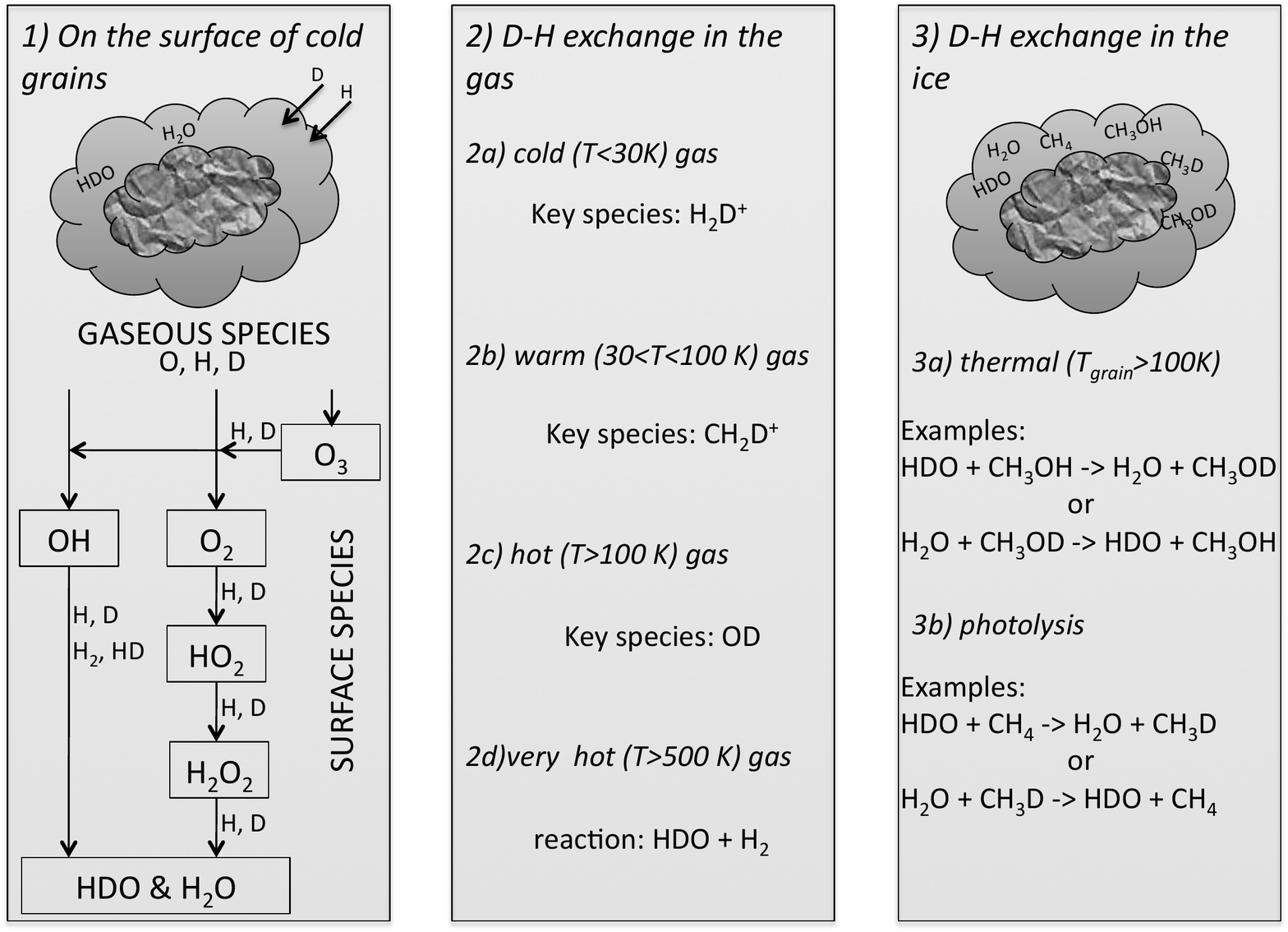}
 \caption{\small The three principal routes of water formation and
   deuteration. {\it Left panel:} gas species as atomic O, H and D
   land on the grain surfaces and form water through OH, O$_2$ and
   O$_3$ hydrogenation. {\it Central panel:} in the gas, water
   deuteration proceeds via reactions with H$_2$D$^+$, CH$_2$D$^+$, OD
   and HDO with H$_2$, depending on the gas temperature. {\it Right
     panel:} D and H atoms can also be exchanged between frozen
   species and water while they are on the grain surfaces.}
  \label{fig:D-WATER}
  \end{center}
\end{figure*}

Therefore, generally speaking, the basic molecule for the deuterium
fractionation is H$_2$D$^+$ (and D$_2$H$^+$ and D$_3^+$ in extreme
conditions). The cause for the enhancement of H$_2$D$^+$ with respect
to H$_3^+$ and, consequently, deuterium fractionation is the larger
mass (equivalent to a higher zero energy) of H$_2$D$^+$ with respect
to H$_3^+$, which causes the activation barrier in step 2. The
quantity which governs whether the barrier can be overcome and,
consequently, the deuterium fractionation is the temperature: the
lower the temperature the larger the deuterium fractionation.

Besides, if abundant neutrals and important destruction partners of
H$_3^+$ isotopologues, such as O and CO, deplete from the gas-phase
(for example because of the freeze-out onto dust grains in cold and
dense regions; \S \ref{sec:the-pre-stell} and \S
\ref{sec:the-prot-disk}), the deuterium fraction further increases
\citep{1984ApJ...287L..47D,2003ApJ...591L..41R}. This
is due to the fact that the destruction rates of all the H$_3^+$
isotopologues drop, while the formation rate of the deuterated species
increases because of the enhanced H$_3^+$ abundance.

There is another factor that strongly affects the deuterium
fractionation: the ortho-to-para abundance ratio of H$_2$
molecules. In fact, if this ratio is larger than $\sim10^{-3}$, the
internal energy of the ortho H$_2$ molecules (whose lowest energy
level is $\sim175$ K) can be enough to overcome the H$_2$D$^+$ + H$_2$
$\rightarrow$ HD + H$_3^+$ barrier and limit the H$_2$D$^+$/H$_3^+$
ratio
\citep{2002P&SS...50.1287G,2006A&A...449..621F,2013JPCA..117.9800R}. In
general, it is believed that ortho and para H$_2$ are formed on the
surface of dust grains with a statistical ratio of
3:1. Proton-exchange reactions then convert ortho- into para- H$_2$,
especially at the low temperatures of dense cloud cores, where the
ortho-to-para H$_2$ ratio is predicted to drop below 10$^{-3}$
\citep{2013A&A...554A..92S}.

So far we have discussed the deuterium fractionation routes in cold ($T\la
40-50$ K) gas. Different routes occur in warm ($30\la T \la 100$ K)
and hot ($100\la T \la 1000$ K) gas. In warm gas ($T \la 70-80$~K),
the D-atoms can be transferred to molecules by CH$_2$D$^+$,
whose activation barrier of the reaction with H$_2$ is larger than
that of H$_2$D$^+$ \citep{2000A&A...361..388R}.
At even higher temperatures, OD transfers D-atoms from HD to water
molecules \citep{2010MNRAS.407..232T}. In these last two cases,
CH$_2$D$^+$ and OD play the role of H$_2$D$^+$ at lower
temperatures. At $\ga 500$ K water can directly exchange D
and H atoms with H$_2$
\citep{1981A&A....93..189G,1994GeCoA..58.2927L}.

Finally, some molecules, notably water, are synthesised on the
surfaces of interstellar and interplanetary grains by addition and/or
substitution of H and D atoms (\S \ref{sec:fract-water}). In this
case, the deuterium fractionation depends on the D/H ratio of the
atomic gas. As discussed previously in this section, the
enhanced abundance of H$_2$D$^+$ (D$_2$H$^+$ and D$_3^+$) also implies
an increased atomic D/H ratio in the gas ({\it Step 3b} in Fig.
\ref{fig:D-chemistry-scheme}), as deuterium atoms are formed
upon dissociative recombination of the H$_3^+$ deuterated
isotopologues, whereas H atoms maintain an about constant density of
$\ga$1\,cm$^{-3}$, determined by the balance between surface formation
and cosmic-ray dissociation of H$_2$ molecules
\citep{2003ApJ...585..823L}.

\subsection{\textbf{Deuterium fractionation of water}}\label{sec:fract-water}

Given the particular role of deuterated water in understanding the
Solar System history, we summarize the three major processes (reported
in the literature) that cause the water deuteration. They are
schematically shown in Fig. \ref{fig:D-WATER}.
\\
{\it 1) Formation and deuteration on the surfaces of cold grains:} in
cold molecular clouds and star forming regions, water is mostly formed
by H and D atoms addition to O, O$_2$ and O$_3$ on the grain surfaces,
as demonstrated by several laboratory experiments
\citep{1998ApJ...498..710H,2010A&A...512A..30D,2012ApJ...749...67O}. In
this case, therefore, the key parameter governing the water
deuteration is the atomic D/H ratio in the gas, which depends on the
H$_2$D$^+$/H$_3^+$ ratio as discussed in the previous section.
\\
\begin{table*}
 \begin{tabular}{| l | l | l |}    
   \hline
   {\bf Acronym} & \multicolumn{2}{|l|}{\bf Definition} \\
   \hline
   $aa$    & \multicolumn{2}{|l|}{Amino acids.} \\
   A$_V$  & \multicolumn{2}{|l|}{Visual extinction, measured in
     magnitudes (mag), proportional to the H nuclei column density in the line of sight.} \\
   ALMA  & \multicolumn{2}{|l|}{Atacama Large Millimeter Array.}\\
   CSO   & \multicolumn{2}{|l|}{Caltech Submillimeter Observatory.}\\
   HSO  &  \multicolumn{2}{|l|}{Herschel Space Observatory.}\\
   JCMT & \multicolumn{2}{|l|}{James Clerk Maxwell Telescope.}\\
   IOM  & \multicolumn{2}{|l|}{Insoluble Organic Matter.} \\
   ISM  & \multicolumn{2}{|l|}{Inter-Stellar Medium.} \\
   NOEMA   & \multicolumn{2}{|l|}{Northern Extended Millimeter Array.}\\
   PSN  & \multicolumn{2}{|l|}{Proto Solar Nebula.}\\
   SOC  & \multicolumn{2}{|l|}{Soluble Organic Compounds (same as SOM).} \\
   SOM  & \multicolumn{2}{|l|}{Soluble Organic Matter (same as SOC).} \\
   VLT &  \multicolumn{2}{|l|}{Very Large Telescope.}\\
   \hline \hline
   \multicolumn{3}{|l|}{\bf Astrophysical key objects
     mentioned in the chapter}\\
   \hline
   PSC   & \multicolumn{2}{|l|}{Pre-stellar Cores: dense and cold
     condensations, they are the first step of
     solar type star formation.} \\
   Class 0 & \multicolumn{2}{|l|}{Class 0 sources: 
     youngest known solar type protostars.}\\ 
   Hot corino & \multicolumn{2}{|l|}{The warm and dense inner regions
     of the envelopes of Class 0 protostars.} \\
   PPD  & \multicolumn{2}{|l|}{Proto-Planetary Disks: disks of
     material surrounding young protostars.}\\
   \hline \hline
   \multicolumn{3}{|l|}{\bf Origin of the Solar System bodies
     mentioned in the chapter}\\
   \hline
   JFC   &  \multicolumn{2}{|l|}{Jupiter-family comets: ecliptic
     short-period comets, whose reservoir is the Kuiper belt.}\\
   &  \multicolumn{2}{|l|}{~~~Probably formed in
     the trans-Neptunian region.}\\
   OCC & \multicolumn{2}{|l|}{Oort-cloud comets:  
     long-period comets, whose reservoir is the Oort  cloud.}\\
   &  \multicolumn{2}{|l|}{~~~Probably formed in the Uranus-Neptune
     region, with some contribution from the }\\
   &  \multicolumn{2}{|l|}{~~~Jupiter-Saturn region.}\\
   CCs    & \multicolumn{2}{|l|}{Carbonaceous Chondrites: the most
     primitive meteorites, mostly coming from the main
     belt (2-4 AU).} \\
   IDPs  & \multicolumn{2}{|l|}{Interplanetary Dust Particles: At least 50\% are fragments of comets,
     the rest fragments of main belt}\\
   & \multicolumn{2}{|l|}{~~~asteroids.}\\
   \hline \hline
   \multicolumn{3}{|l|}{\bf Measure of molecular deuteration:}\\
   \hline
   \multicolumn{3}{|l|}{1) In geochemical and biochemical studies, the
     molecular deuteration is measured by the deuterium enrichment in a sample}\\
   \multicolumn{3}{|l|}{~~~~compared to that of a chosen
     standard value, in \textperthousand ~with respect to the standard ratio.}\\
   \multicolumn{3}{|l|}{~~~~For deuterium: $\delta {\rm D}
     \textperthousand =
     \frac{D/H_{sample}-D/H_{VSMOW}}{D/H_{VSMOW}}\times 10^3$} \\
   \multicolumn{3}{|l|}{2) In astrochemical studies, the
     molecular deuteration is measured by the heavy/light isotope abundance.}\\
   \multicolumn{3}{|l|}{3) In the PSN models, the molecular deuteration is given by the enrichment factor $f$, defined as the ratio between the molecule}\\ 
   \multicolumn{3}{|l|}{~~~~D/H and the PSN D/H.}\\
   \hline
   {\bf Acronym} & {\bf Definition} & {\bf D/H~~~~~~~~~~~~~$\delta$D ~~~ $f$~~~~  references}\\
   \hline
   $A_D$ & Cosmic elemental deuterium abundance &
   $1.6\times10^{-5}$ ~-900 ~~0.8~~~ a\\
   PSN D/H & Deuterium abundance in the PSN & $2.1\times10^{-5}$
   ~-860 ~~1.0~~~ b\\
   VSMOW & Vienna Standard Mean Ocean Water (refers to evaporated
   ocean waters) & $1.5\times10^{-4}$ ~~~~~~0 ~~~7.1~~~~c\\
    \hline
  \end{tabular}
  \caption{\small  Definitions of symbols and terms used in the
    Chapter. References: a: \cite{2007SSRv..130..367L}. b: \cite{1998SSRv...84..239G}. c: \cite{Lecuyer1998}.}
  \label{tab:definitions}
  \end{table*}
  {\it 2) Hydrogen-Deuterium exchange in the gas phase:} as for any
  other molecule, D-atoms can be transferred from H$_2$D$^+$
  (D$_2$H$^+$ and D$_3^+$) to H$_2$O in cold gas (Sec.
  \ref{sec:the-chem-proc}), and more efficiently through the HD +
  OH$^+$ $\rightarrow$ HDO + H and HD + OH$_2^+$ $\rightarrow$ HDO +
  H$_2$ reactions. In warm gas, it is in principle possible to have
  direct exchange between HD and H$_2$O to form HDO
  \citep{1981A&A....93..189G,1994GeCoA..58.2927L}. However, being a
  neutral-neutral reaction, it possesses an activation barrier
  \citep{1977AREPS...5...65R}, which makes this route very slow at
  $T\la500$ K. On the contrary, for temperatures high enough ($\geq
  100$ K), the OH + HD and OD + H$_2$ reactions can form HDO. Based on
  modelling, \cite{2010MNRAS.407..232T} demonstrated that the HD + O
  $\rightarrow$ OD + H followed by the OD + H$_2$ $\rightarrow$ HDO +
  H reaction is indeed a major route for the HDO formation in warm
  gas.
  \\
\begin{figure*}[bt]
  \begin{center}
    \includegraphics[width=14cm]{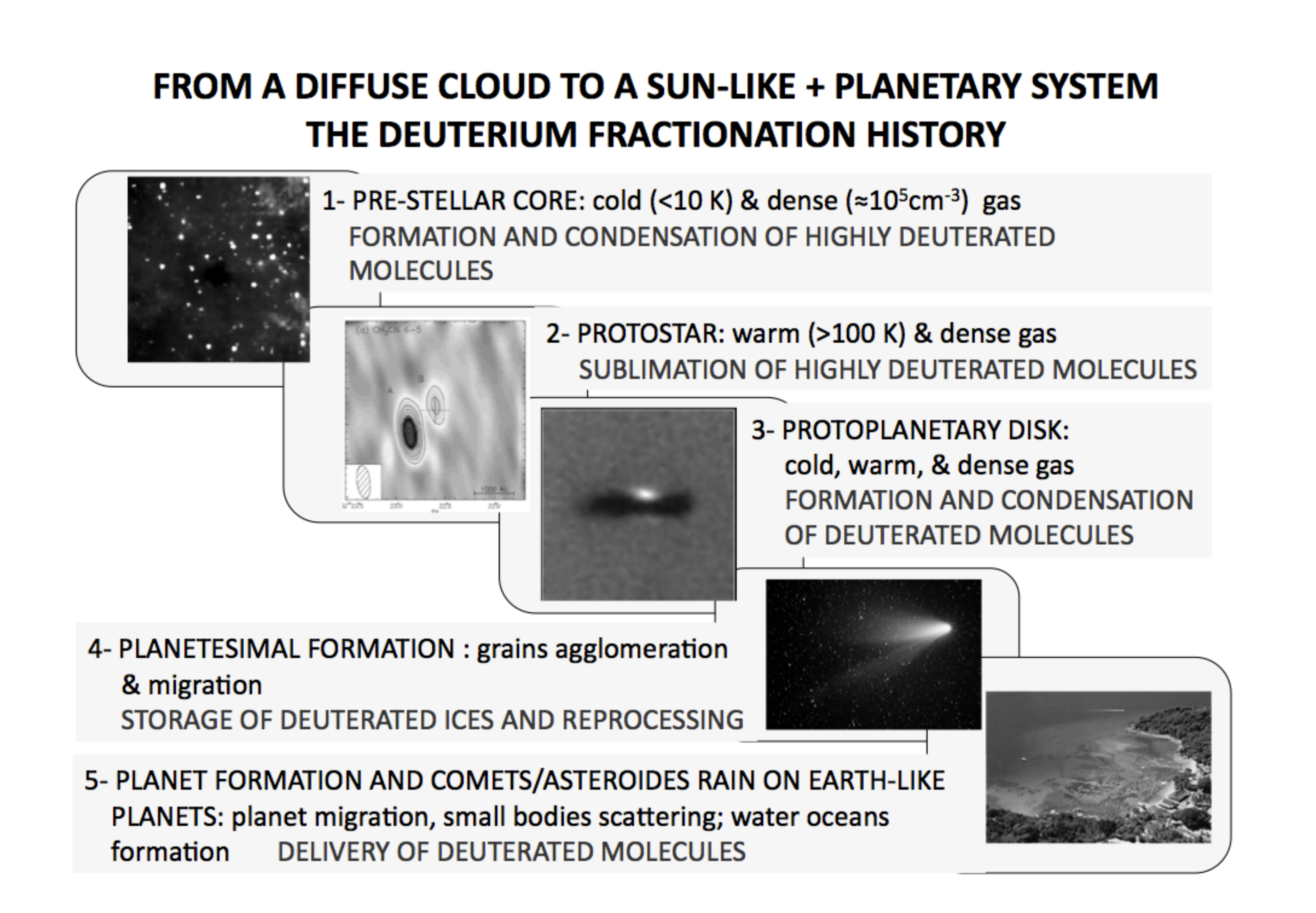}
    \caption{\small Schematic summary of the different phases that likely
      gave birth to the Solar System, with highlighted the main deuterium
      processes (adapted from Caselli \& Ceccarelli 2012).}
    \label{fig:sec3-fig1}
  \end{center}
\end{figure*}
{\it 3) Isotopic exchange between solid H$_2$O and HDO with other
  solid species:} laboratory experiments have shown that D and H atoms
can be exchanged between water ice and other molecules trapped in the
ice, like for example CH$_3$OH
\citep{2009A&A...496L..21R,2011ApJ...738..133G}. Very likely, the
exchange occurs during the ice sublimation phase, with the
re-organisation of the crystal. Similarly, H-D exchange in ice can be
promoted by photolysis \citep{2009ApJ...703.1030W}.  Note that this mechanism
not only can alter the HDO/H$_2$O abundance ratio in the ice, but also
it can pass D-atoms to organic matter trapped in the ice, enriching it
of deuterium.

\subsection{\textbf{Towards a common language}}\label{sec:language}

This chapter has the ambition to bring together researchers from
different communities. One of the disavantages, which we aim to
overcome here, is that these different communities do not always speak
the same language. Table \ref{tab:definitions} is a a sort of
dictionary which will help the reader to translate the chapter in
her/his own language.  In addition, several acronyms used throughout
this chapter are also listed in the table. With this, we are ready now
to start our voyage through the different objects.

\section{\textbf{A BRIEF HISTORY OF THE SOLAR SYSTEM FORMATION}}\label{sec:brief-hist}

According to the widely accepted scenario, the five major phases of
solar type star formation are (Fig. \ref{fig:sec3-fig1}):
\begin{itemize}
\item [1: ] {\bf Pre-stellar cores.}  These are the starting point of
  Solar-type star formation. In these "small clouds" with evidence of
  contraction motions, contrarily to starless cores, matter slowly
  accumulates toward the center, causing the increase of the density
  while the temperature is kept low ($\la$10\,K). Atoms and molecules
  in the gas-phase freeze-out onto the cold surfaces of the sub-micron
  dust grains, forming the so called icy grain mantles. This is the
  moment when the deuterium fractionation is most effective: the
  frozen molecules, including water, acquire a high deuterium
  fraction.
\item [2: ]{\bf Protostars. } The collapse starts, the
  gravitational energy is converted into radiation and the envelope
  around the central object, the future star, warms up. When and where
  the temperature reaches the mantle sublimation temperature ($\sim$
  100--120 K), in the so-called hot corinos, the molecules in
  the mantles are injected into the gas-phase, where they are 
  observed via their rotational lines. Complex organic molecules,
  precursors of prebiotic species, are also detected at this stage.
\item [3: ] {\bf Protoplanetary disks. } The envelope dissipates with
  time and eventually only a circumstellar, protoplanetary disk
  remains. In the hot regions, close to the central object or the disk
  surface, some molecules, notably water, can be D-enriched via
  neutral-neutral reactions. In the cold regions, in the midplane,
  where the vast majority of matter resides, the molecules formed in
  the protostellar phase freeze-out again onto the grain mantles,
  where part of the ice from the pre-stellar phase may still be
  present. The deuterium fractionation process becomes again important.
\item [4: ] {\bf Planetesimals formation. } The process of
  "conservation and heritage" begins. The sub-micron dust grains
  coagulate into larger rocks, called planetesimals, the seeds of the
  future planets, comets and asteroids. Some of the icy grain mantles
  are likely preserved while the grains glue together. At least part
  of the previous chemical history may be conserved in the building
  blocks of the forming planetary system rocky bodies and eventually
  passed as an heritage to the planets. However, migration and
  diffusion may scramble the original distribution of the D-enriched
  material.
\item [5: ] {\bf Planet formation.}  The last phase of rocky planet
  formation is characterized by giant impacts between planet embryos,
  which, in the case of the Solar System, resulted in the formation of
  the Moon and Earth. Giant planets may migrate, inducing a scattering
  of the small bodies all over the protoplanetary disk. Oceans are
  formed on the young Earth and, maybe in other rocky planets. The
  leftovers of the process become comets and asteroids. In the Solar
  System, their fragments continuously rain on Earth releasing the
  heritage stored in the primitive D-enriched ices. Life takes over
  sometime around 2 billion years after the Earth and Moon formation
  \citep{2010NatGe...3..522C}.
\end{itemize}  
In the rest of the chapter, we will discuss each of these steps, the
measured deuterium fractionation and the processes responsible for
that.

\section{\textbf{THE PRE-STELLAR CORE PHASE}}\label{sec:the-pre-stell}

\subsection{The structure of pre-stellar cores}

\begin{figure*}[bt]
  \begin{center}
    \includegraphics[width=14cm]{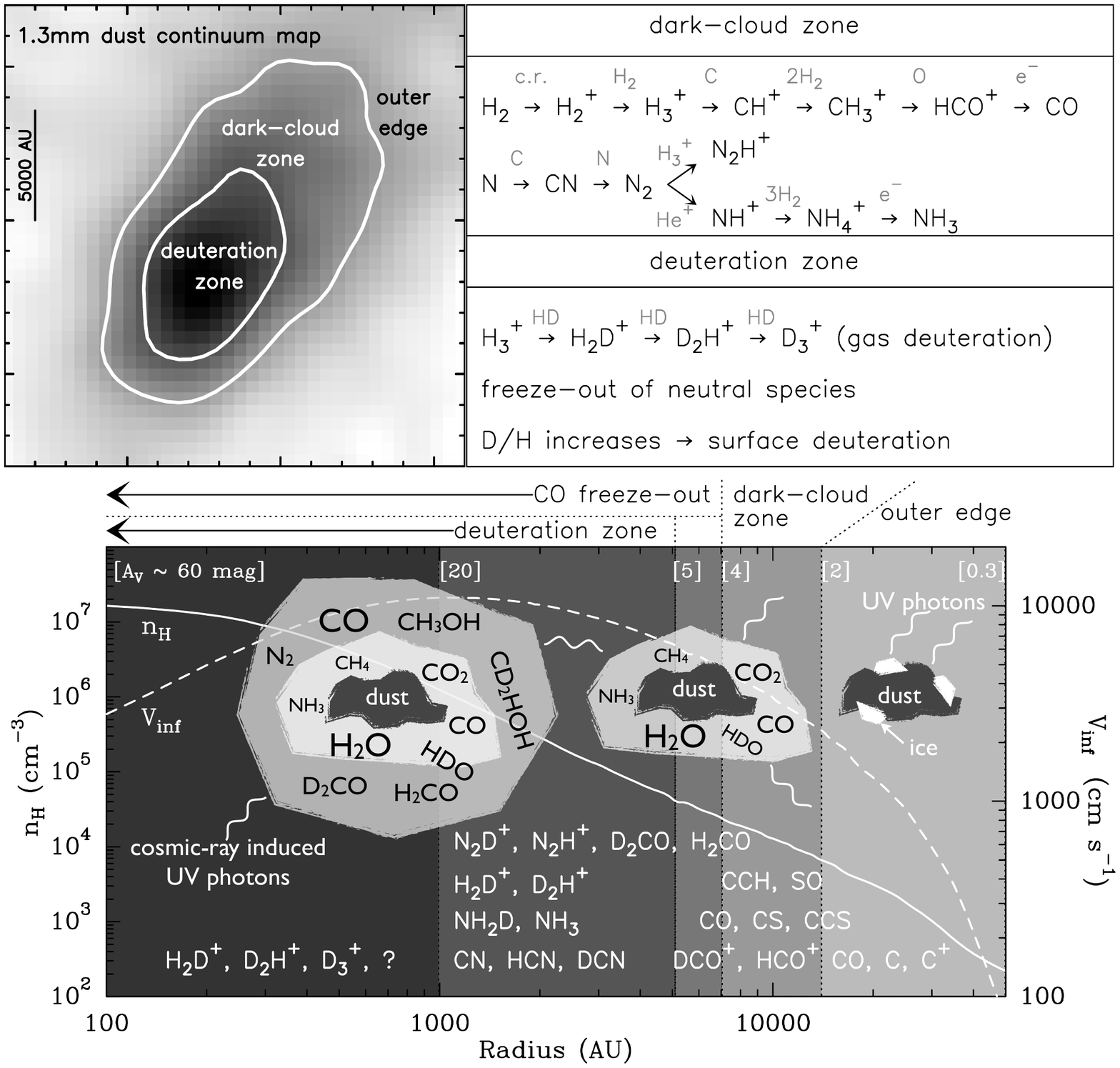}
    \caption{\small Schematic summary of the main physical and
      chemical characteristics of a pre-stellar core. {\it Upper left
        panel.} The grey scale is the 1.2\,mm dust continuum emission
      of L1544, the prototypical pre-stellar core \citep[data
      from][]{1999MNRAS.305..143W}. White contours mark the transition
      from the outer edge to the dense core and the deuteration zone,
      where CO is highly frozen onto dust grains. {\it Upper right
        panel.} Main chemical processes in the dark-cloud zone and the
      deuteration zone. Grey symbols above the arrows are the reaction
      partners and the arrow indicates the reaction direction. {\it
        Bottom panel.} Radial slice of L1544, indicating the number
      density ($n_{\rm H}$), infall velocity ($V_{\rm inf}$) profiles,
      the visual extinction A$_V$, the main molecular diagnostic tools
      in the various zones and a rough summary of dust grain evolution
      within the pre-stellar core \citep[see
      also][]{2012A&ARv..20...56C}.}
    \label{fig:psc-fig1}
  \end{center}
\end{figure*}
Stars form within fragments of molecular clouds, the so-called dense
cores \citep[][]{1989ApJS...71...89B}, produced by the combined
action of gravity, magnetic fields and turbulence
\citep{2007ARA&A..45..565M}. Some of the starless dense cores can be
transient entities and diffuse back into the parent cloud, while
others (the pre-stellar cores) will dynamically evolve until the
formation of one or more planetary systems. It is therefore important
to gather kinematics information to identify pre-stellar cores, which
represent the initial conditions in the process of star and planet
formation. Their structure and physical characteristics depend on the
density and temperature of the surrounding cloud, i.e. on the external
pressure (Tan et al., this volume).  The well-studied pre-stellar
cores in nearby molecular clouds have sizes of $\simeq$10,000\,AU,
similar to the Oort cloud, masses of a few Solar masses and visual
extinctions $\ga$50\,mag.  They are centrally concentrated
\citep{1999MNRAS.305..143W}, with central densities larger than 10$^5$
H$_2$ cm$^{-3}$\citep{2003ApJ...585L..55B,2005ApJ...619..379C,
  2008ApJ...683..238K}, central temperatures close to 7\,K
\citep{2007A&A...470..221C, 2007A&A...467..179P} and with evidence of
subsonic gravitational contraction \citep{2010MNRAS.402.1625K} as well
as gas accretion from the surrounding cloud
\citep{2011ApJ...734...60L}.  The ESA Herschel satellite has detected
water vapour for the first time toward a pre-stellar core and unveiled
contraction motions in the central $\simeq$1,000\,AU
\citep{2012ApJ...759L..37C}: large amounts of water (a few Jupiter
masses, mainly in ice form) are transported toward the future
Solar-type star and its potential planetary system.

A schematic summary of the main chemical and physical characteristics
of pre-stellar cores is shown in Figure~\ref{fig:psc-fig1}. The upper
left panel shows one of the
best studied objects: L1544 in the Taurus Molecular Cloud Complex,
140\,pc away. The largest white contour roughly indicates the size of
the dense core and the outer edge represents the transition region
between L1544 and the surrounding molecular cloud, where the
extinction drops below $\simeq$4\,mag and photochemistry becomes
important. This is where water ice copiously form on the surface of
dust grains \citep[][]{2007ApJ...655..332W, 2009ApJ...690.1497H} and
low-levels of water deuteration are taking place
\citep{2011ApJ...741L..34C, 2012ApJ...748L...3T}.  Within the {\it
  dark-cloud zone}, where the carbon is mostly locked in CO, gas-phase
chemistry is regulated by ion-molecule reactions
(Fig.\,\ref{fig:psc-fig1}, top right panel).

Within the central $\simeq$7,000\,AU, the volume density becomes
higher than a few $\times$10$^4$\,cm$^{-3}$ (see bottom panel) and the
freeze-out timescale ($\simeq$10$^9$/$n_{\rm H}$\,yr) becomes shorter
than a few $\times$10$^4$\,yr. This is the {\it deuteration zone},
where the freeze-out of abundant neutrals such as CO and O, the main
destruction partners of H$_3^+$ isotopologues, favour the formation of
deuterated molecules (see Sect.\,~\ref{sec:the-chem-proc}). Deuteration is one of the
main chemical process at work, making deuterated species the best
tools to study the physical structure and dynamics of the earliest
phases of star formation.

\subsection{Deuterium fractionation}

Pre-stellar cores chemically stand out among other starless dense
cores, as they show the largest deuterium fractions \citep[$>$10\%;
][]{2003ApJ...585L..55B, 2005ApJ...619..379C, 2007A&A...467..179P} and
CO depletions \citep[$>$90\% of CO molecules are frozen onto dust
grain surfaces;][]{1999ApJ...523L.165C, 2002A&A...389L...6B}. The
largest D-fractions have been observed in N$_2$H$^+$
\citep[N$_2$D$^+$/N$_2$H$^+$ $\simeq$
0.1--0.7;][]{2005ApJ...619..379C, 2007A&A...467..179P}, ammonia
\citep[NH$_2$D/NH$_3$ $\simeq$ 0.1--0.4;][]{2001ApJ...554..933S,
  2007A&A...470..221C}, and formaldehyde \citep[D$_2$CO/H$_2$CO
$\simeq$ 0.01--0.1;][]{2003ApJ...585L..55B}. HCN, HNC and HCO$^+$ show
somewhat lower deuterations \citep[between 0.01 and
0.1;][]{2001ApJS..136..579T, 2003ApJ...594..859H, 1995ApJ...448..207B,
  2002ApJ...565..344C}. \citet{2013ApJ...769L..19S} recently detected
doubly deuterated cyclopropenylidene (c-C$_3$D$_2$), finding a
c-C$_3$D$_2$/c-C$_3$H$_2$ abundance ratio of about
0.02. Unfortunately, no measurement of water deuteration is available
yet.

Pre-stellar cores are the strongest emitters of the ground state
rotational transition of ortho-H$_2$D$^+$ \citep{2003A&A...403L..37C}
and the only objects where para-D$_2$H$^+$ has been detected
\citep{2004ApJ...606L.127V, 2011A&A...526A..31P}. It is interesting to
note that the strength of the ortho-H$_2$D$^+$ line does not correlate
with the amount of deuterium fraction found in other molecules.  This
is probably due to variations of the H$_2$D$^+$ ortho-to-para ratio in
different environments \citep{2008A&A...492..703C}, an important clue
in the investigation of how external conditions affect the chemical
and physical properties of pre-stellar cores \citep[see
also][]{2013ApJ...765...59F}.

\subsection{Origin of deuterium fractionation}

Pre-stellar cores are "deuterium fractionation factories". The reason
for this is twofold: firstly, they are very cold (with typical gas and
dust temperatures between 7 and 13\,K). This implies a one-way
direction of the reaction H$_3^+$ + HD $\rightarrow$ H$_2$D$^+$ +
H$_2$, the starting point of the whole process of molecular
deuteration (see Fig.\ref{fig:D-chemistry-scheme}, step 2 and
3a). Secondly, a large fraction of neutral heavy species such as CO
and O freeze-out onto dust grains.
As mentioned in \S \ref{sec:the-chem-proc}, the disappearance of
neutrals from the gas phase implies less destruction events for
H$_3^+$ and its deuterated forms, with consequent increase of not just
H$_2$D$^+$ but also D$_2$H$^+$ and D$_3^+$ \citep{1984ApJ...287L..47D,
  2003ApJ...591L..41R, 2004A&A...418.1035W}. This simple combination
of low temperatures and the tendency for molecules to stick on icy
mantles on top of dust grains, can easily explain the observed
deuterium fraction measured in pre-stellar cores \citep[see
also][]{2005A&A...438..585R}.

The case of formaldehyde (H$_2$CO) deuteration requires an extra note,
as not all the data can be explained by gas-phase models including
freeze-out.  As discussed in \citet{2003ApJ...585L..55B}, another
source of deuteration is needed. A promising mechanism is the chemical
processing of icy mantles (surface chemistry), coupled with partial
desorption of surface molecules upon formation
\citep{2006FaDi..133...51G}. In particular, once CO freezes out onto
the surface of dust grains, it can either be stored in the ice
mantles, or be "attacked" by reactive elements, in particular atomic
hydrogen.
In the latter case, CO is first transformed into HCO, then
formaldehyde and eventually into methanol (CH$_3$OH). In pre-stellar
cores, deuterium atoms are also abundant because of the dissociative
recombination of the abundant H$_2$D$^+$, D$_2$H$^+$ and possibly
D$_3^+$ (see Fig.\ref{fig:D-chemistry-scheme}, step 3b). Chemical
models predict D/H between 0.3 and 0.9 in the inner zones of
pre-stellar cores with large CO freeze-out \citep{2003ApJ...591L..41R,
  2012ApJ...748L...3T}, implying a large deuteration of formaldehyde
and methanol on the surface of dust grains (see dust grain cartoons
overlaid on the bottom panel of Fig.\,\ref{fig:psc-fig1}). Thus, the
measured large deuteration of gas-phase formaldehyde in pre-stellar
cores (and possibly methanol, although this still awaits for
observational evidence) can be better understood with the contribution
of surface chemistry, as a fraction of surface molecules can desorb
upon formation (thanks to their formation energy).

\section{\textbf{THE PROTOSTAR PHASE}}\label{sec:prot-phase}

\subsection{\textbf{The structure of the Class 0 protostars}}
Class 0 sources are the youngest protostars. Their luminosity is
powered by the gravitational energy, namely the material falling
towards the central object, accreting it at a rate of $\la 10^{-5}$
M$_\odot$/yr. They last for a short period, $\sim 10^5$ yr
\citep{2009ApJS..181..321E} (see also Dunham et al. this volume).  The
central object, the future star, is totally obscured by the collapsing
envelope, whose sizes are $\sim 10^4$ AU, as the pre-stellar cores (\S
\ref{sec:the-pre-stell}). It is not clear whether a disk 
exists at this stage \citep{2010A&A...512A..40M}, as the original
magnetic field frozen on the infalling matter tends to inhibit its
formation (Z.-Y. Li et al., this volume). On the contrary, powerful
outflows of supersonic matter are one of the notable characteristics
of these objects (Frank et al., this volume).

In Class 0 protostars, the density of the envelope increases with
decreasing distance from the centre ($n\propto r^{-3/2}$), as well as
the temperature \citep{1996ApJ...471..400C,2002A&A...389..908J}. From
a chemical point of view, the envelope is approximatively divided in
two regions, delimitated by the ice sublimation temperature (100--120
K): a cold outer envelope, where molecules are more or less frozen
onto the grain mantles, and an inner envelope, called hot corino,
where the mantles, built up during the pre-stellar core phase (\S
\ref{sec:the-pre-stell}), sublimate \citep{2000A&A...357L...9C}
\citep[see][for a more accurate description of the Class 0
protostars]{2012A&ARv..20...56C}.  This transitions occurs at distances
from the center between 10 and 100 AU, depending on the source
luminosity \citep{2004A&A...416..577M,2005A&A...442..527M}. Relevant
to this chapter, in the hot corinos, the species formed at the
pre-stellar core epoch are injected into the gas phase, bringing
memory of their origin. For different reasons, the outer envelope and
the hot corino have molecules highly enriched in deuterium: in the
first case because of the low temperatures and CO depletion (\S
\ref{sec:the-chem-proc} and \S \ref{sec:the-pre-stell}), in the second
case because of the inheritance of the pre-stellar ices.

\subsection{\textbf{Deuterium fractionation}}
Class 0 protostars are the objects where the highest deuterium
fractionation has been detected so far and the first where the extreme
deuteration, called in literature super-deuteration, has been
discovered \citep{1998A&A...338L..43C,2007prpl.conf...47C}: doubly and
even triply deuterated forms with D/H enhancements with respect to the
elemental D/H abundance (Tab. \ref{tab:definitions}) of up to 13
orders of magnitude.

\begin{figure}[tb]
 \includegraphics[width=8cm]{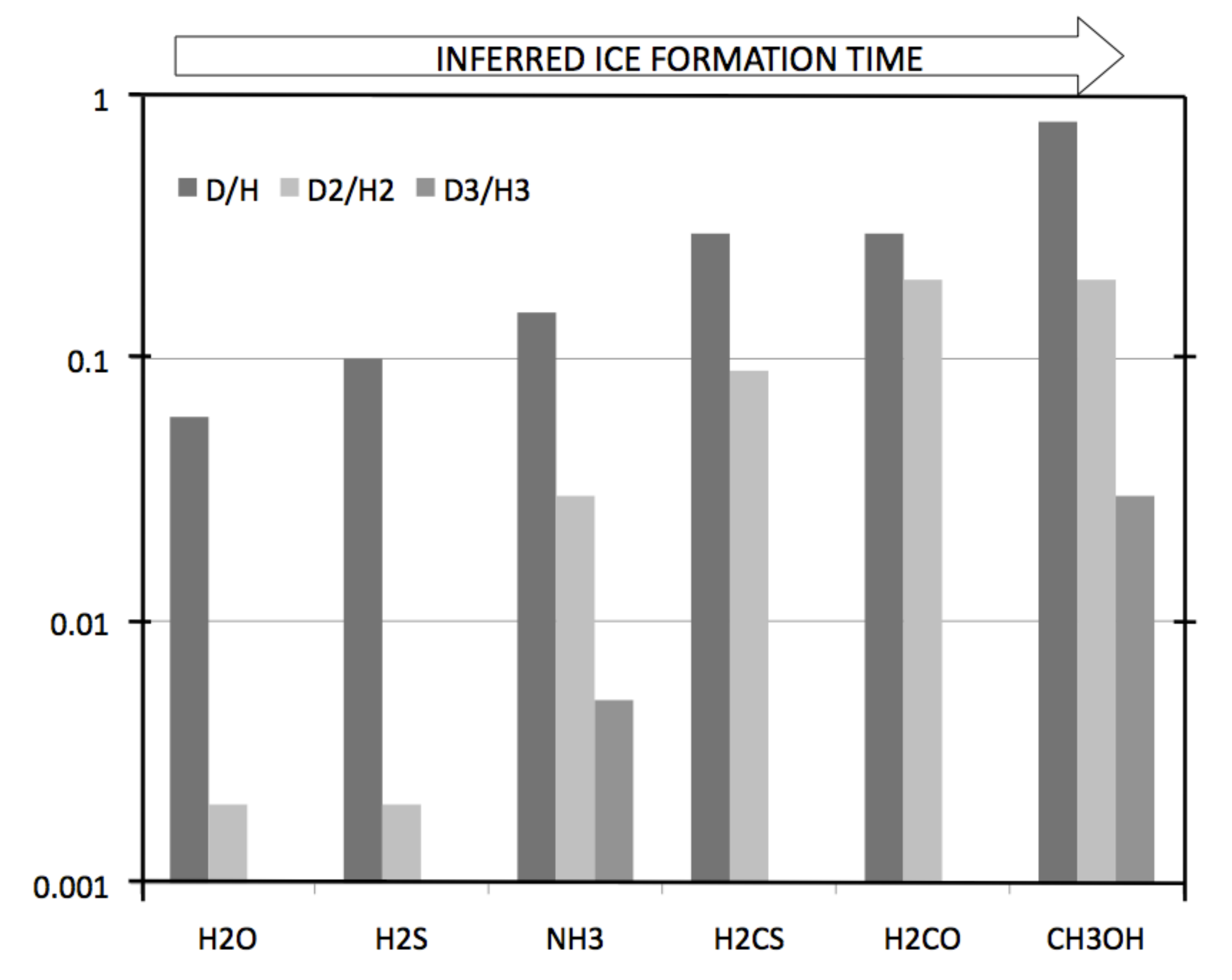}
 \caption{\small Measured deuteration ratios of singly, doubly and
   triply deuterated isotopologues \cite[adapted
   from][]{2012A&ARv..20...56C}. Modelling the formation of H$_2$O,
   H$_2$CO and CH$_3$OH
   \citep{2011ApJ...741L..34C,2013A&A...550A.127T} suggests that the
   increasing deuteration reflects the formation time of the species
   on the ices. References: H$_2$O: \cite{2011A&A...527A..19L},
   \cite{2012A&A...539A.132C}, \cite{2013ApJ...768L..29T},
   \cite{2007ApJ...659L.137B}, \cite{2010A&A...521L..31V}; H$_2$S:
   \cite{2003ApJ...593L..97V}; NH$_3$:
   \cite{2001ApJ...552L.163L},\cite{2002A&A...388L..53V}; H$_2$CS:
   \cite{2005ApJ...620..308M}; H$_2$CO: \cite{1998A&A...338L..43C};
   \cite{2006A&A...453..949P}; CH$_3$OH:
   \cite{2002A&A...393L..49P,2004A&A...416..159P,2006A&A...453..949P}. }
\label{fig:protostar-1}
\end{figure}

The first and the vast majority of measurements were obtained with
single dish observations, so that they cannot disentangle the outer
envelope and the hot corino, if not indirectly by modelling the line
emission in some cases
\citep{2012A&A...539A.132C,2013A&A...560A..39C}. The following species
with more than two atoms of deuterium have been detected (see
\cite{2007prpl.conf...47C} for a list of singly deuterated species):
formaldehyde
\citep[D$_2$CO:][]{1998A&A...338L..43C,2006A&A...453..949P}, methanol
\citep[CHD$_2$OH and
CD$_3$OH:][]{2002A&A...393L..49P,2004A&A...416..159P,2006A&A...453..949P},
ammonia \citep[NHD$_2$ and
ND$_3$:][]{2001ApJ...552L.163L,2002A&A...388L..53V,2002ApJ...571L..55L},
hydrogen sulphide \citep[D$_2$S:][]{2004ApJ...606L.127V},
thioformaldehyde \citep[D$_2$CS: ][]{2005ApJ...620..308M} and water
\citep[D$_2$O:][]{2007ApJ...659L.137B,2010A&A...521L..31V}.  In a few
cases, interferometric observations provided us with measurements of
water deuterium fractionation in the hot corinos
\citep{2010A&A...522L...1C,2012A&A...541A..39P,2013ApJ...768L..29T}. Finally,
recent observations have detected deuterated species in molecular
outflows \citep{2010A&A...522L...1C,2012ApJ...757L...9C}. The
situation is graphically summarized in Fig. \ref{fig:protostar-1}.

We note that the methanol deuteration depends on the bond energy of
the functional group which the hydrogen is located. In fact, the
abundance ratio CH$_2$DOH/CH$_3$OD is larger than $\sim20$
\citep{2006A&A...453..949P,2011A&A...528L..13R}, whereas it should be
3 if the D-atoms were statistically distributed. To explain this, it
has been invoked that CH$_3$OD could be selectively destroyed in the
gas-phase \citep{1997ApJ...482L.203C}, that H--D exchange reactions in
solid state could contribute to reduce CH$_3$OD
\citep{2005ApJ...624L..29N}, or, finally, that the CH$_3$OD under
abundance is due to D--H exchange with water in the solid state
\citep{2009A&A...496L..21R,2011A&A...528L..13R}. The reason for this
over-deuteration of the methyl group with respect to the hydroxyl
group may help to understand the origin of the deuterium fractionation
in the different functional groups of the Insoluble Organic Matter (\S
\ref{sec:meteorites}; see Fig. \ref{fig:iom-fig3}). This point will be
discussed further in \S \ref{sec:summaryD}.

\begin{figure}[tb]
 \includegraphics[width=8cm]{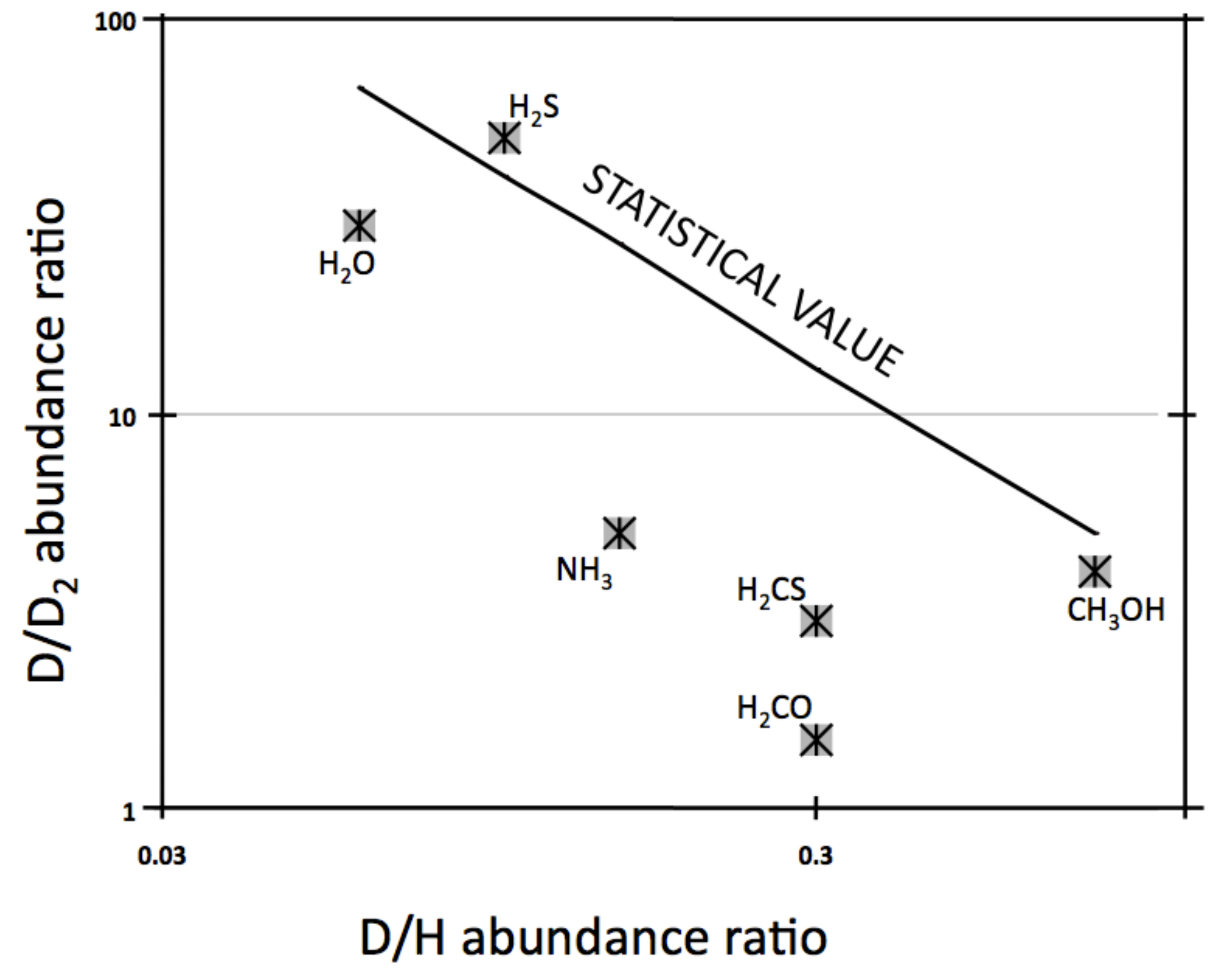}
 \caption{\small Measured ratios of singly to doubly deuterated
   isotopologues of the species marked in the plot. The line shows the
   statistical value if the D atoms were statistically distributed in
   the molecules formed on the grain surfaces. References as in
   Fig. \ref{fig:protostar-1}. Adapted from \cite{2012A&ARv..20...56C}.}
\label{fig:protostar-2}
\end{figure}

The measure of the abundance of doubly deuterated species can also
help to understand the formation/destruction routes of the
species. Fig. \ref{fig:protostar-2} shows the D/D$_2$ abundance ratio
of the molecules in Fig. \ref{fig:protostar-1}. For species forming on
the grain surfaces, if the D atoms were purely statistically
distributed, namely just proportional to the D/H ratio, then it would
hold: D-species/D$_2$-species = 4 (D-species/H-species)$^{-1}$. As
shown in Fig. \ref{fig:protostar-2}, this is not the case for H$_2$O,
NH$_3$, H$_2$CS and H$_2$CO. A plausible explanation is that the D and
D$_2$-bearing forms of these species are formed at different times on
the grain surfaces: the larger the deuterium fraction the younger the
species (see \S 5.3).

In the context of this chapter, the deuteration of water deserves
particular attention. Being water very difficult to observe with
ground based telescopes (van Dishoeck et al., this volume),
measurements of HDO/H$_2$O exist only towards four Class 0 sources,
and they are, unfortunately, even in disagrement. Table
\ref{tab:protostars} summarizes the situation.
\begin{table}[bt]
  \centering
  \begin{tabular}{|l|c|r|r|}
    \hline
    Source & HDO/H$_2$O & Note & Ref. \\
               & ($10^{-3}$) & & \\ \hline
    IRAS16293-2423 & 3--15 & out & 1\\
                             & 4--51 & hc & 1\\
                             & 0.7--1.2 & hc & 2\\
    NGC1333-IRAS2A & 9--180 & out & 3\\
                               & $\sim 1$ & hc & 3, 4\\
                               & 3--80    & hc & 5\\
    NGC1333-IRAS4A & 5--30    & hc & 5\\
    NGC1333-IRAS4B & $\leq 0.6$ & hc & 6\\
\hline
  \end{tabular}
  \caption{\small Measurement of the HDO/H$_2$O ratio in Class 0
    sources. In the third column we report whether the measure refers
    to the outer envelope (out) or the hot corino (hc). References: 1: \cite{2012A&A...539A.132C,2013A&A...553A..75C}. 2: \cite{2012A&A...541A..39P}. 3: \cite{2011A&A...527A..19L}. 4: \cite{2013ApJ...769...19V}. 5: \cite{2013ApJ...768L..29T}. 6:
    \cite{2010ApJ...725L.172J}.}
\label{tab:protostars}
\end{table}
With the ALMA facility it will be possible to settle the issue in a
near future.  Nonetheless, one thing is already clear from those
measurements: the deuteration of water is smaller than that of the
other molecules. This is confirmed also by the upper limits on the
HDO/H$_2$O in the solid phase: $\la1$\%
\citep{2003A&A...399.1009D,2003A&A...410..897P}.

\begin{figure*}[bt]
  \begin{center}
    \includegraphics[width=13cm]{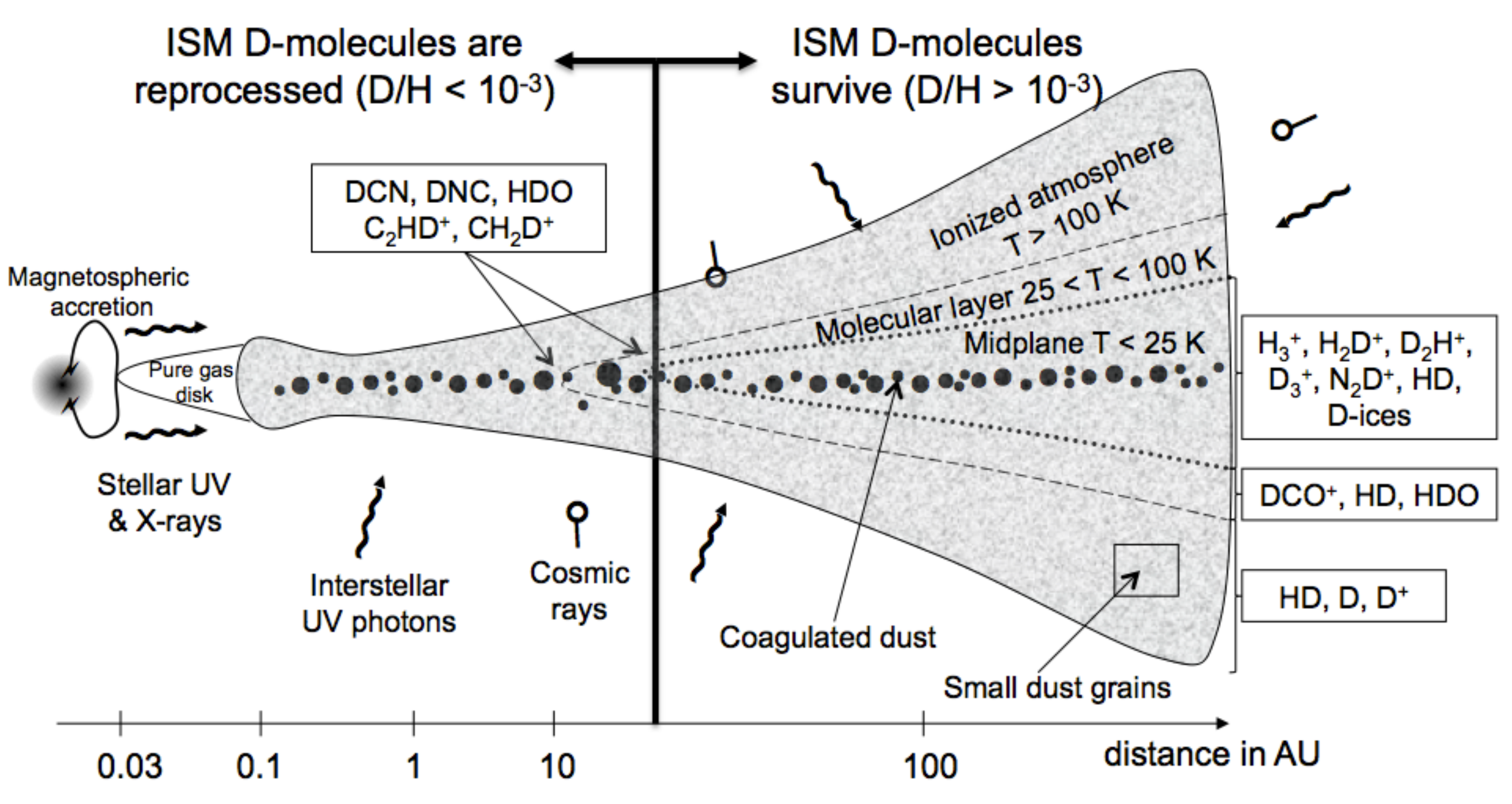}  
    \caption{\small  Schematic vertical cut through a protoplanetary disk
      around a Sun-like T Tauri star. From the chemical point of view,
      the disk can be divided onto 3 layers: (1) the cold midplane
      dominated by ices and grain surface processes, (2) the warmer
      molecular layer with active gas-grain processes, where many
      gaseous molecules concentrate, and (3) the hot irradiated
      atmosphere dominated by photochemistry and
      populated by ions and small molecules. 
}
  \label{fig:disk-fig1}
 \end{center}
 \end{figure*}

\subsection{\textbf{Origin of the deuterium fractionation}}
As discussed in \S \ref{sec:the-chem-proc} and \S
\ref{sec:the-pre-stell}, the key elements for a large deuterium
fractionation are a cold and CO depleted gas. The large D-enrichment
observed in Class 0 sources, and especially in the hot corinos is,
therefore, necessarily inherited from the pre-stellar core phase. The
short lifetimes estimated for the Class 0 sources support that
statement.
The different D fractionation then tells us of the past history of the
Class 0 sources. Following \S \ref{sec:the-pre-stell}, it is very
likely that water, formaldehyde and methanol were formed on the grain
surface at that epoch, by hydrogenation of O and CO. The O
hydrogenation leading to water occurs first, during the molecular
cloud phase, and, once formed, water remains frozen on the grains
\citep{2009ApJ...690.1497H}. CO hydrogenation leading to H$_2$CO and,
subsequently, to CH$_3$OH occurs later, in the pre-stellar core phase
\citep{2012A&A...538A..42T}. Consequently, the water D enrichment is
lower than that of H$_2$CO and CH$_3$OH. In other words, the
increasing of the D-enrichment, linked to the H$_2$D$^+$/H$_3^+$ ratio
(\S \ref{sec:the-chem-proc}) corresponds to a later formation of the
molecule. Models by \cite{2011ApJ...741L..34C} and
\cite{2012ApJ...748L...3T,2013A&A...550A.127T} provide quantitative
estimates in excellent agreement with the observed values.
The same explanation applied to Fig. \ref{fig:protostar-2}, tells us
that H$_2$CO was synthesised in a large interval of time, so that the
singly and doubly deuterated forms did not inherit the same D/H atomic
ratio. On the contrary, CH$_3$OH was formed on a short time
interval. Again, this agrees with model predictions


\section{\textbf{THE PROTOPLANETARY DISK PHASE}}\label{sec:the-prot-disk}

\subsection{The structure of protoplanetary disks}
The protostellar phase has a short life time ($\sim$0.5\,Myr), during
which the powerful outflows disperse the surrounding 
envelope, reducing the infall of material from the envelope onto the
young disks and the accretion of material from the disk to the
protostar. The central few hundred astronomical units are eventually
revealed and the "classical" protoplanetary disk (PPD) phase starts. This is
the last, dynamically active stage, lasting several Myr, when the
remaining $\sim 0.001-0.1M_\odot$ of gas and dust is accreted onto the
newly formed central star and assembled into a planetary system.  

A summary of the physical and chemical structure of a protoplanetary
disk of a Proto Solar Nebula (PSN) analogue (a T Tauri star) is
sketched in Fig.\ref{fig:disk-fig1} (a more detailed description can
be found in Dutrey et al. and Testi et al., this volume; \cite[see
also][and references
therein]{2012A&ARv..20...56C,2013ChRv..113.9016H}. Starting from the
location of the accreting protostar, undergoing magnetospheric
accretion and emitting UV photons and X-rays, we note that the first
fraction astronomical unit is mainly deprived of dust, because of the
large temperatures, above the dust melting point
($\simeq$1500\,K). The dusty disk starts at around 0.1\,AU with an
inner wall, puffed-up by the high temperature
\citep{2002A&A...389..464D}. The rest of the disk, up to a few hundred
AU, can be approximately divided in three layers: (1) the ionised
atmosphere, where the chemistry is dominated by photodissociation due
to stellar and interstellar UV photons as well as stellar X-rays
\citep[][]{2010A&A...518L.125T, 2012A&A...547A..69A}. (2) A warm
molecular layer, where molecules survive but photochemistry still
plays an important role
\citep[][]{2005ApJ...635L..85D,2010ApJ...714.1511H}. (3) The cooler
zones around the midplane, where the temperature drops to values close
to those measured in pre-stellar cores (\S
\ref{sec:the-pre-stell}). The densities (up to $\sim
10^{11}$\,cm$^{-3}$) are significantly higher than those of
pre-stellar cores (up to a few times 10$^{7}$\,cm$^{-3}$), so that
dust coagulation, freeze-out and deuterium fractionation are expected
to proceed faster \citep{2005A&A...440..583C}.  The temperature
increases vertically, away from the midplane, and radially, toward the
central object. For a T Tauri star as that in
Fig.\,\ref{fig:disk-fig1}, the inner region (1$<$ radius $<$ 30\,AU)
has temperatures between $\simeq$30 and 150\,K, while lower
temperatures are found in the midplane at $\ga$ 30\,AU.

The sketch in Fig. \ref{fig:disk-fig1} is of course only a rough
snapshot.  During the several Myr of its life time, the disk thermal
and density structure, X-ray/UV radiation intensities, and grain
properties undergo enormous transformation (as traced by infrared
spectroscopy, (sub-)millimeter/centimeter interferometry and advanced
disk modelling \citep[][]{2007ApJ...659..705A, Bouwman_ea08,
  2009ApJ...700.1502A, Guilloteau_ea11a, 2011ARA&A..49...67W,
  2013ApJ...762...48G}. The grain growth and sedimentation toward the
midplane leads to more transparent and thus hotter disk atmosphere and
flatter vertical structure, shrinking the zone when gas and dust are
collisionally coupled and have similar temperatures.  The disk
chemical composition changes along with the evolution of the physical
conditions \citep{Aikawa_ea06, Nomura_ea07a, Cleeves_ea11a,
  Ilee_ea11a, Fogel_ea11, Vasyunin2011, ANDES}. As a result of the
increasing X-ray/UV-irradiation with time, the molecularly-rich, warm
layer shifts closer to the cold midplane and the pace of surface
chemistry can be delayed by the grain growth.

\subsection{Deuterium fractionation}

Observational studies of the detailed chemical structure of
protoplanetary disks require high sensitivity and high angular
resolution measurements, which started to become available only in the
past few years. For this reason, measurements of deuterium fractions
are very sparse and so far limited to two PSN analogue disks (the T
Tauri disks TW Hya and DM Tau) and one disk around a 2\,M$_{\odot}$
star (the Herbig Ae star HD~163296; \cite{2013arXiv1307.3420M}, not
discussed here).

{\bf TW Hya.} The first deuterated molecule, DCO$^+$, was detected by
\citet{2003A&A...400L...1V} in the T Tauri star TW Hya. The
DCO$^+$/HCO$^+$ abundance ratio was found $\simeq$0.04, similar to
that measured toward pre-stellar cores (\S \ref{sec:the-pre-stell}).
The first image of the DCO$^+$ emission has been obtained by
\citet{Qi_ea08} using the Submillimeter Array (SMA). They found
variations of the DCO$^+$/HCO$^+$ across the disk, increasing outward
from $\sim$1\% to $\sim$10\% and with a rapid falloff at radii
$\ga$90\,AU.  DCN/HCN was also measured in the same observing run and
found to be $\simeq$2\%. Combining SMA with ALMA data,
\citet{2012ApJ...749..162O} have inferred that DCO$^+$ and DCN do not
trace the same regions: DCN is centrally concentrated and traces inner
zones. This is in agreement with the theoretical predictions that DCN
can also be produced in gas warmer than 30\,K via the intermediate
molecular ion CH$_2$D$^+$ (see \S\ref{sec:the-chem-proc}), unlike
DCO$^+$ which is linked to H$_2$D$^+$.  The CO snow line has been
recently imaged using ALMA observations of N$_2$H$^+$ (a molecular ion
known to increase in abundance when CO starts to freeze-out, as CO is
one of the destruction partners), finding a radius of $\sim$30\,AU
\citep{2013arXiv1307.7439Q}. Deuterated species are expected to thrive
close to the CO snow line and more observations should be planned to
improve our understanding of the deuterium fractionation around this
disk.

{\bf DM~Tau.} \cite{2006A&A...448L...5G} found a DCO$^+$/HCO$^+$
abundance ratio toward DM~Tau ($\simeq$0.004) about one order of
magnitude lower than toward TW Hya, indicating interesting differences
between apparently similar sources, which need to be explored in more
detail.

\subsection{Origin of the deuterium fractionation}

The deuterium fractionation in protoplanetary disks is expected to
occur mainly in the cold ($T \la$30\,K) midplane region, where the CO
freeze-out implies large abundances of the H$_3^+$ deuterated
isotopologues \citep{2005A&A...440..583C}. The fractionation route
based on CH$_2$D$^+$ is also effective in the warm inner disk midplane
and molecular layer (Fig.\ref{fig:disk-fig1} and \S
\ref{sec:set-stage}). Modern models of protoplanetary disks include a
multitude of deuterium fractionation reactions and multiply-deuterated
species, still however without nuclear spin-state processes.

\citet{Willacy_07} have investigated deuterium chemistry in the outer
disk regions.
They found that the observed DCO$^+$/HCO$^+$ ratios can
be reproduced, but, in general, the deuteration of gaseous molecules
(in particular doubly deuteration) can be greatly changed by chemical
processing in the disk. In their later study, \citet{Willacy_Woods09}
have investigated deuterium chemistry in the inner 30~AU, using the
same physical structure and accounting for gas and dust thermal
balance.  While a good agreement between the model predictions and
observations of several non-deuterated gaseous species in a number of
protoplanetary disks was obtained, the calculated D/H ratios for ices
were higher than measured in the Solar System comets. 

\citet{2010MNRAS.407..232T} have focused on understanding deuterium
fractionation in the inner warm disk regions and investigated the
water D/H ratio in dense ($\ga 10^6$~cm$^{-3}$) and warm ($\sim
100-1000$~K) gas, caused by photochemistry and neutral-neutral
reactions. Using the T~Tau disk structure calculated with ``ProDiMo''
\citep{Woitke_ea09}, they predicted that in the terrestrial
planet-forming region at $\la 3$~AU the water D/H ratio may reach
$\sim 1\%$, which is higher than the value of $\approx 1.5\times
10^{-4}$ measured in the Earth ocean water
(Tab. \ref{tab:definitions}). As mentioned in \S \ref{sec:fract-water},
the formation of HDO at high temperatures ($\ga$ 100\,K) proceeds
through the OH + HD $\rightarrow$ OD + H$_2$ reaction (see
Fig.\ref{fig:D-WATER}).

As presented above, our knowledge on deuterium chemistry in
protoplanetary disks, particularly from the observational perspective,
is still rather poor. Only the deuterated molecules DCN and DCO$^+$
have been detected so far \citep[plus the recent Herschel detection of
HD toward TW~Hya;][]{2013Natur.493..644B}.  The analysis of their
relatively poorly resolved spectra shows that DCO$^+$ emission tends
to peak toward outer colder disk regions, while DCN emission appears
to be more centrally peaked. Their chemical differentiation seems to
be qualitatively well-understood from the theoretical point of view,
with DCN being produced at warmer temperatures ($\la 70-80$~K) by
fractionation via CH$_2$D$^+$ and C$_2$HD$^+$, and DCO$^+$ being
produced at cold temperatures ($\la 10-30$~K) by fractionation via
H$_3^+$ isotopologues. However, theoretical models of the inner warm
disk regions still tend to overpredict the D/H ratios of molecules
observed in comets.  ALMA, soon in full capabilities, will provide us
with a wealth of new precious data on deuterated species in many
protoplanetary disks, which will shed light on the chemical and
physical processes during this delicate phase of planet formation.

\section{\textbf{COMETS}}\label{sec:the-comets}

\subsection{\textbf{The origin of comets}}
Having retained and preserved pristine material from the Proto Solar
Nebula (PSN) at the moment of their accretion, comets contain unique
clues to the history and evolution of the Solar System. Their study
provides the natural link between interstellar matter and Solar System
bodies and their formation. Comets accreted far from the Sun, where
ices could condense, and have remained for most of their lifetime
outside the orbit of Pluto. According to the reference scenario, the
Oort cloud, which is the reservoir of long-period comets (also called OCC), was
populated by objects gravitationally scattered from the Uranus-Neptune
region, with some contribution from the Jupiter-Saturn region
\citep{Dones2004,Brasser2008}. The reservoir of ecliptic short-period
comets (the so-called Jupiter-family comets JFC) is the Kuiper belt
beyond Neptune, more precisely the scattered disk component of that
population characterized by highly eccentric orbits
\citep{1997Sci...276.1670D}. In the current picture described by the
Nice model, the present scattered disk is the remnant of a much larger
scattered population produced by the interaction of the primordial
planetesimal disk with a migrating Neptune. In the simulations of
\citet{Brasser2013} made in the framework of the Nice model, the Oort
cloud and the scattered disk formed simultaneously from objects from
the same parent population.
In summary, either the two comet reservoirs
come from the same population of objects formed in the Uranus-Neptune
zone, or Oort-cloud comets formed on average closer to the Sun.
Finally, a third option reported in the literature is that Oort-cloud
comets were captured from other stars when the Sun was in its birth
cluster \citep{Levison2010}.  The molecular and isotopic composition
in these two reservoirs can provide key answers in this respect.
\begin{figure}[bt]
\includegraphics[angle = 0, width = 8.2cm]{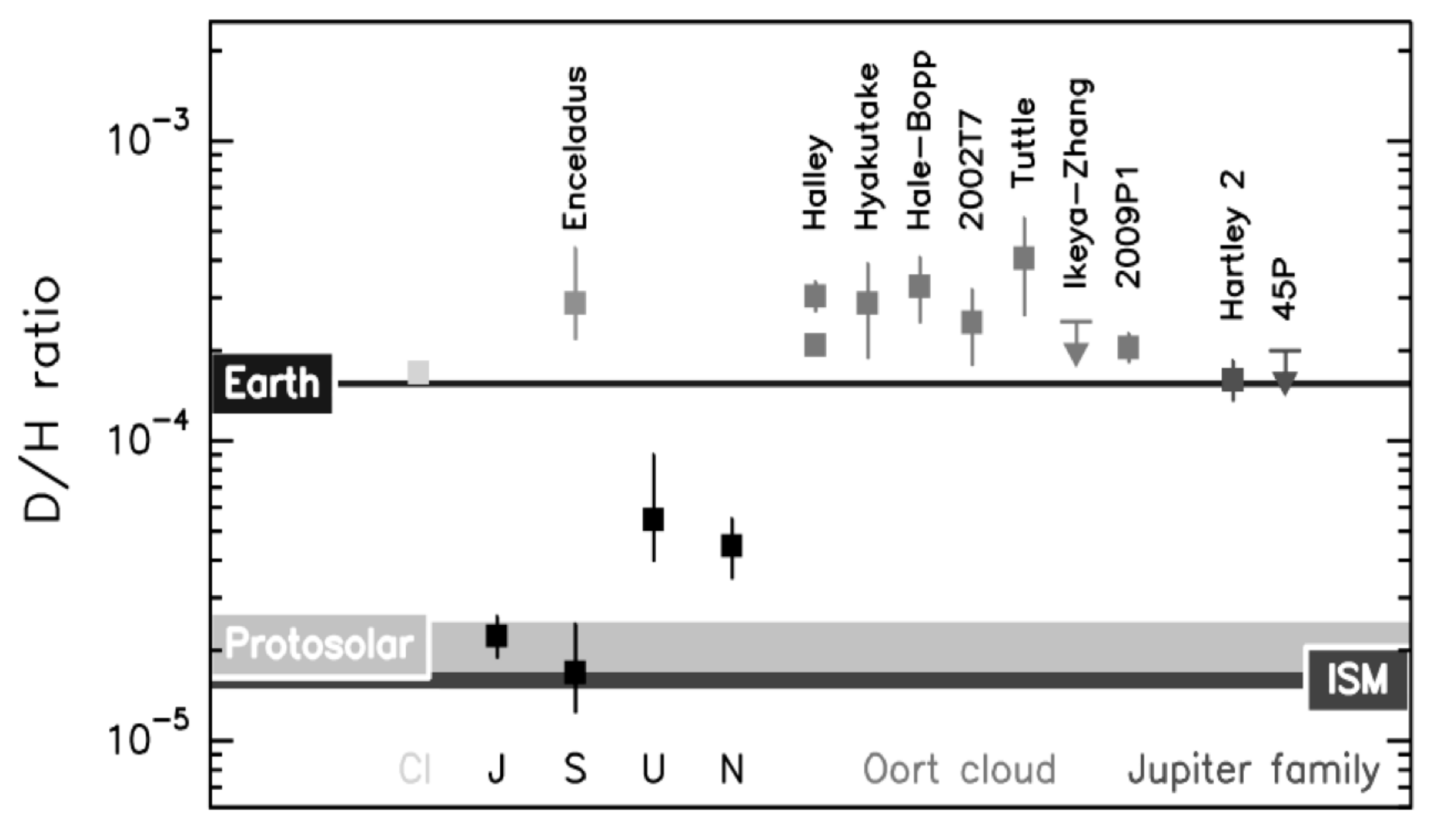}
\caption{\small D/H ratio in the water of comets compared to values in
  carbonaceous meteorites (CI), Earth's oceans (VSMOW; see Tab. 1),
  and Enceladus. Displayed data for planets, the interstellar medium,
  and the PSN refer to the value in H$_2$ (see text for
  details). Adapted from \citet{Lis13}.}
\label{fig:Dcomets}
\end{figure}
\subsection{\textbf{Deuterium fractionation}}
Because of the faint signatures of deuterated species, data were only
obtained in a handful of bright comets. The first measurements of the
D/H ratio in cometary H$_2$O were obtained in comet 1P/Halley from
mass-resolved ion-spectra of H$_3$O$^+$ acquired with the Ion Mass
Spectrometer \citep{Bal95} and the Neutral Mass Spectrometer
\citep{Ebe95} intruments onboard the European Giotto spacecraft. These
independent data provided precise D/H values of (3.08
$^{+0.38}_{-0.53}$) $\times~10^{-4}$ and (3.06 $\pm$ 0.34)
$\times~10^{-4}$ \citep{Bal95,Ebe95}. We note however that
\citet{Bro12} reexamined the mass-spectrometer measurements,
reevaluating these values to 2.1 $\times~10^{-4}$.

From observations undertaken with CSO and JCMT, HDO was
detected in the bright long-period comets C/1996 B2 (Hyakutake) and
C/1995 O1 (Hale-Bopp) from its \mbox{1$_{01}$-0$_{00}$} line at
464.925~GHz \citep{Boc98,Mei98a}. The derived D/H values ((2.9 $\pm$
1.0) $\times~10^{-4}$ and (3.3 $\pm$ 0.8) $\times~10^{-4}$, for comets
Hyakutake and Hale-Bopp, respectively) are in agreement with the
determinations in comet Halley (Fig.~\ref{fig:Dcomets}).  Finally,
observations of the HDO 1$_{10}$--1$_{01}$ transition at 509.292 GHz
in the Halley-type comet 153P/Ikeya-Zhang yielded D/H $<$ 2.5
$\times~10^{-4}$ \citep{Biv06}.

Using the new high resolution spectrograph CRIRES of the ESO VLT,
\citet{Vil09} observed the HDO ro-vibrational
transitions near 3.7 $\mu$m in the Halley-family comet 8P/Tuttle
originating from the Oort cloud.  Twenty three lines were co-added to
get a marginal detection of HDO, from which a formal value of D/H of
(4.09 $\pm$ 1.45) $\times~10^{-4}$ was derived.

In cometary atmospheres, water photodissociates into mainly OH and
H. The OD/OH ratio was measured to be (2.5 $\pm$ 0.7) $\times~10^{-4}$
in the Oort-Cloud comet C/2002 T7 (LINEAR) through ground-based
observations of the OH A $^2$$\Sigma$$^+$ -- X $^2$$\Pi$$_i$
ultraviolet bands at 310 nm obtained with VLT feeding the UVES
spectrograph \citep{Hut08}. No individual OD line was detected, but a
marginal 3$\sigma$ detection of OD was obtained by co-adding the
brightest lines. Atomic deuterium (D) emission was also discovered
during ultraviolet observations of comet C/2001 Q4 (NEAT) in April
2004 using the spectrograph STIS of the Hubble Space Telescope
\citep{Wea08}.  The Lyman-$\alpha$ emission from both D and atomic
hydrogen were detected, from which a preliminary value D/H = (4.6
$\pm$ 1.4) $\times~10^{-4}$ was derived, assuming that H$_2$O is the
dominant source of the observed D and H, as is likely the case. The
strength of the optical method is that both normal and rare
isotopologues have lines in the same spectral interval and are
observed simultaneously, avoiding problems related to comet variable
activity.

The most recent D/H measurements in cometary water were acquired using
the ESA Herschel satellite. The HDO 1$_{10}$--1$_{01}$ transition at
509.292 GHz was observed using the HIFI spectrometer in the
Jupiter-family comets 103P/Hartley 2 and
45P/Honda-Mrkos-Pajdu\v{s}\'{a}kov\'{a} \citep{Har11,Lis13}, and in
the Oort-cloud comet C/2009 P1 (Garradd) \citep{Boc12}.  Observations
of HDO were interleaved with observations of the H$_2$O and
H$_2^{18}$O 1$_{10}$--1$_{01}$ lines. Since the H$_2$O ground state
rotational lines in comets are optically thick, optically thin lines
of H$_2^{18}$O provide, in principle, a more reliable reference for
the D/H determination. The HDO/H$_2^{18}$O was measured to be 0.161
$\pm$ 0.017 for comet Hartley 2, i.e., consistent with the VSMOW
(0.1554 $\pm$ 0.0001). The HDO/H$_2^{18}$O value of 0.215 $\pm$
0.023 for comet Garradd suggests a significant difference (3-$\sigma$)
in deuterium content between the two comets.  \citet{Har11} derived a
D/H ratio of (1.61 $\pm$ 0.24) $\times~10^{-4}$ for comet Hartley,
assuming an H$_2$$^{16}$O/H$_2$$^{18}$O ratio of 500 $\pm$ 50, which
encompasses the VSMOW value and values measured in cometary water
\citep{Jeh09}. For comet Garradd, the derived D/H ratio is (2.06 $\pm$
0.22) $\times~10^{-4}$ based on the HDO/H$_2^{16}$O production rate
ratio, and (2.15 $\pm$ 0.32) $\times~10^{-4}$, using the same method
as \citet{Har11}. Herschel observations in the Jupiter family comet
45P/Honda-Mrkos-Pajdu\v{s}\'{a}kov\'{a} were unsuccessful in detecting
the HDO 509 GHz line, but resulted in a sensitive 3$\sigma$ upper
limit for the D/H ratio of $2.0 \times 10^{-4}$ which is consistent
with the value measured in comet 103P/Hartley 2 and excludes the
canonical pre-Herschel value measured in Oort-cloud comets of $\sim$ 3
$\times~10^{-4}$ at the 4.5$\sigma$ level \citep{Lis13}.

Besides H$_2$O, the D/H ratio was also measured in HCN, from the
detection of the J=5--4 rotational transition of DCN in comet
Hale-Bopp with the JCMT \citep{Mei98b}. The inferred D/H value in HCN
is 2.3 $\times~10^{-3}$, i.e., seven times the value measured in water
in the same comet. Finally, D/H upper limits for several molecules were
obtained, but most of these upper limits exceed a few percent
\citep{Cro04}. For CH$_4$, observed in the near-IR region, the most
stringent D/H upper limits are $\sim$ 0.5 \% \citep[][and references
therein]{Gib12}.
\begin{table}[tb]
  \centering
  \begin{tabular}{|l|c|c|}
    \hline
    Comet & D/H & Type\\ \hline
    Halley & $(3.1\pm 0.5)\times 10^{-4}$ & OCC\\
    Hyakutake    & $(2.9\pm 1.0)\times 10^{-4}$ & OCC\\
    Hale-Bopp   & $(3.3\pm 0.8)\times 10^{-4}$ & OCC\\
    2002 T7  & $(2.5\pm 0.7)\times 10^{-4}$ & OCC\\
    Tuttle     & $(4.1\pm 1.5)\times 10^{-4}$ & OCC\\
    Ikeya-Zhang & $\le 2.5\times 10^{-4}$ & OCC\\
    2009 P1 & $(2.06\pm 0.22)\times 10^{-4}$ & OCC\\
    2001 Q4    & $(4.6\pm 1.4)\times 10^{-4}$ & OCC\\
    Hartley 2  & $(1.61\pm 0.24)\times 10^{-4}$ & JFC\\
    45P  & $\le 2.0 \times 10^{-4}$ & JFC\\
    \hline
  \end{tabular}
  \caption{\small Summary of the measured water D/H values in comets (see
    text for the references). The last column report the type of
    comet, Oort-cloud (OCC) or Jupiter-family (JFC).}
  \label{tab:Dcomets}
\end{table}

Table \ref{tab:Dcomets} summarizes the situation and
Fig. \ref{fig:Dcomets} shows the {\it water ice D/H} values measured
so far in comets, compared to the protosolar and the Earth ocean's
value VSMOW (see Tab. \ref{tab:definitions}).
Isotopic diversity is present in the population of Oort-cloud comets,
with members having a deuterium enrichment of up to a factor of two
with respect to the Earth value. The only two Jupiter-family comets
for which the D/H ratio has been measured present low values
consistent with the Earth's ocean value.

\subsection{\textbf{Origin of the deuterium fractionation}}
The high D/H ratio measured in comets cannot be explained by deuterium
exchanges between H$_2$ and H$_2$O in the PSN, and is interpreted as
resulting from the mixing of D-rich water vapor originating from the
pre-solar cloud or outer cold disk, and material reprocessed in the
inner hot PSN (\S \ref{sec:solar-nebula}). The discovery of
an Earth-like D/H ratio in the Jupiter-family comet 103P/Hartley 2
came as a great surprise.  Indeed, Jupiter-family comets were expected
to have D/H ratios higher than or similar to Oort-cloud comets
according to current scenarios \citep{2011ApJ...734L..30K}, implying
that the source regions of these two populations or the models
investigating the gradient of D/H in the PSN should be
revisited. Actually, the only dynamical theory that could explain in
principle an isotopic dichotomy in D/H between Oort-cloud and
Jupiter-family comets is the one arguing that a substantial fraction
of Oort-cloud comets were captured from other stars when the Sun was
in its birth cluster \citep{Levison2010}. Another possible solution is
that there was a large-scale movement of planetesimals between the
inner and outer PSN. According to the Grand Tack scenario
proposed for the early Solar System, when the giant planets were still
embedded in the nebular gas disk, there was a general radial mixing of
the distribution of comets and asteroids \citep{Walsh2011}. Both the
similarity of the D/H ratio in comets 103P/Hartley 2 and
45P/Honda-Mrkos-Pajdu\v{s}\'{a}kov\'{a} with that found in
carbonaceous chondrites, and the isotopic diversity observed in the
Oort-cloud population would be in agreement with this scenario, though
D/H variations in the Jupiter family might be expected as
well. Finally, another possibility is a non monotonic D/H ratio as a
function of the heliocentric distance, as predicted by models
considering a residual infalling envelope surrounding the PSN disk
\citep{2013Icar..290W}. Section \ref{sec:solar-nebula} will explore in
greater detail these models.

\section{\textbf{CARBONACEOUS
    CHONDRITES AND INTERSTELLAR DUST PARTICLES}}\label{sec:meteorites}

Meteorites have fallen to the Earth throughout its
history. Interplanetary Dust Particles (IDPs) are collected in the
stratosphere or in the ices of the Earth polar regions. While most
IDPs are believed to be cometary fragments, probably from
Jupiter-Family comets \citep{2010ApJ...713..816N}, meteorites are
fragments of main-belt asteroids \citep[2-4 AU;][]{2000M&PS...35.1309M},
with the few exceptions of martian and lunar meteorites. Both
meteorites and IDPs provide us with a huge amount of information on
the early phases of the Solar System formation.
Among the various groups of meteorites \cite[see for example the
review by][]{2011IAUS..280..288A}, the carbonaceous chondrites (CCs)
are believed to be the more pristine, little altered by parent bodies,
so that we will focus this section on this group. Briefly, CCs are
stony fragments of primitive asteroids of near solar bulk composition
that were not much heated or differentiated. It is in these mineral
matrixes that clay minerals (the so-called phyllosilicates) and
diverse soluble and insoluble organic materials are found. Based on
the techniques used to study it, the organic matter is classified as
Insoluble Organic Matter (IOM) and Soluble Organic Compounds (SOC;
sometime this is also referred as Soluble Organic Matter or SOM).  The
possible interdependence between minerals and organic materials
through their Solar and pre-Solar formative history is not known and
remains one unexplained feature of chemical evolution. Evidence of CCs
exposure to liquid water also indicates that aqueous processes and
partaking minerals contributed to their final organic inventory.
In this section, we describe the deuterium fractionation in the clays,
IOM, and SOC separately.

\subsection{\textbf{Clays minerals in Carbonaceous Chondrites and Interplanetary Dust Particles}}
\begin{figure}[tb]
 \includegraphics[angle=0,width = 8.2cm]{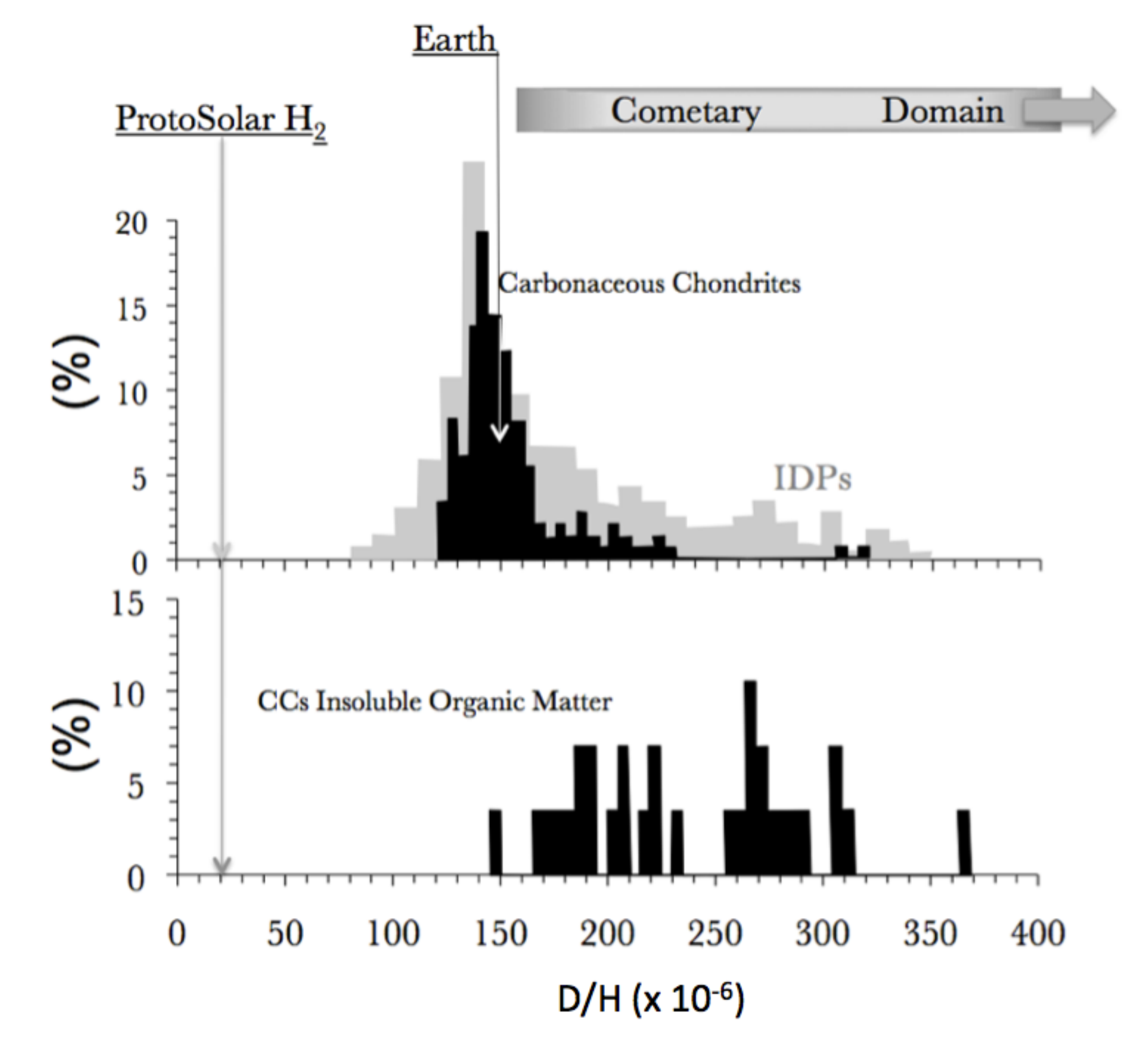}
 \caption{\small Histograms of the D/H ratio in CCs (upper panel,
   black), in IDPs (upper panel, grey) and in the IOM lower
   panel). The figure also shows the PSN and terrestrial water D/H
   ratios (Tab. \ref{tab:definitions}). Note that the D/H ratios
   measured in the IOM ``hot spots'' are not reported here. The figure
   is based on the personal compilation of F. Robert and includes
   published measurements up to June 2013.}
\label{fig:iom-fig1}
\end{figure}
For more than two decades \citep{1979Natur.282..785R}, it has been
known that CC clay minerals exhibit a systematic enrichment in
deuterium relative to the PSN D/H ratio (Tab. \ref{tab:definitions}).
Similarly, clays in IDPs are enriched in deuterium
\citep{1983Natur.305..119Z,2000Natur.404..968M,2000M&PSA..35R..19A}.
Figure \ref{fig:iom-fig1} shows the distribution of the bulk D/H ratio
in the CCs and IDPs. In CCs, the contribution of the D-rich organic
matter (see below) to the variations of the bulk D/H ratio is
negligible relative to the large domain of isotopic variations, i.e.
$\leq 1.5\times 10^{-5}$. However, for IDPs, the relative contribution
of organic H to hydroxyls is not well known so that the tail in the
D/H distribution towards the high values may be caused by the
contribution of D-rich organic hydrogen.  Both CCs and IDPs have D/H
ratios $\sim 7$ times larger than the PSN. Their D/H distribution is
peaked around the value of the terrestrial oceans, the VSMOW
(Tab. \ref{tab:definitions}), with a shoulder extending up to about
3.5 $\times10^{-4}$. The majority of CCs and IDPs possesses,
therefore, a D/H ratio lower than that measured in most comets (\S
\ref{sec:the-comets}).

\cite{2012Sci...337..721A} recently discovered a correlation between
the D/H ratio and the amount of organic hydrogen relative to the bulk
hydrogen (which, as mentioned above, is dominated by the hydroxyl
groups of clays), indicating that the water {\it in the meteorite
  parent body} had a D/H ratio - prior to isotopic exchange with
deuterium-rich organics - as low as $9.5\times 10^{-5}$, namely
smaller than the SWOM value. In addition, these authors found that the
D/H ratio of this pristine water varies according to the meteorite
class, indicating different degree of isotopic enrichment of the PSN
water prior to its introduction in the CCs parent bodies. Based on
these findings, these authors propose that the original D/H ratio in
meteoritic water (i.e. before the clay formaiton) is, therefore,
smaller than what is measured nowadays, with consequences on the
terrestrial water origin (\S \ref{sec:terr-water}).

\subsection{\textbf{Insoluble Organic Matter}}
\begin{figure}[tb]
 \includegraphics[angle=-90,width = 8cm]{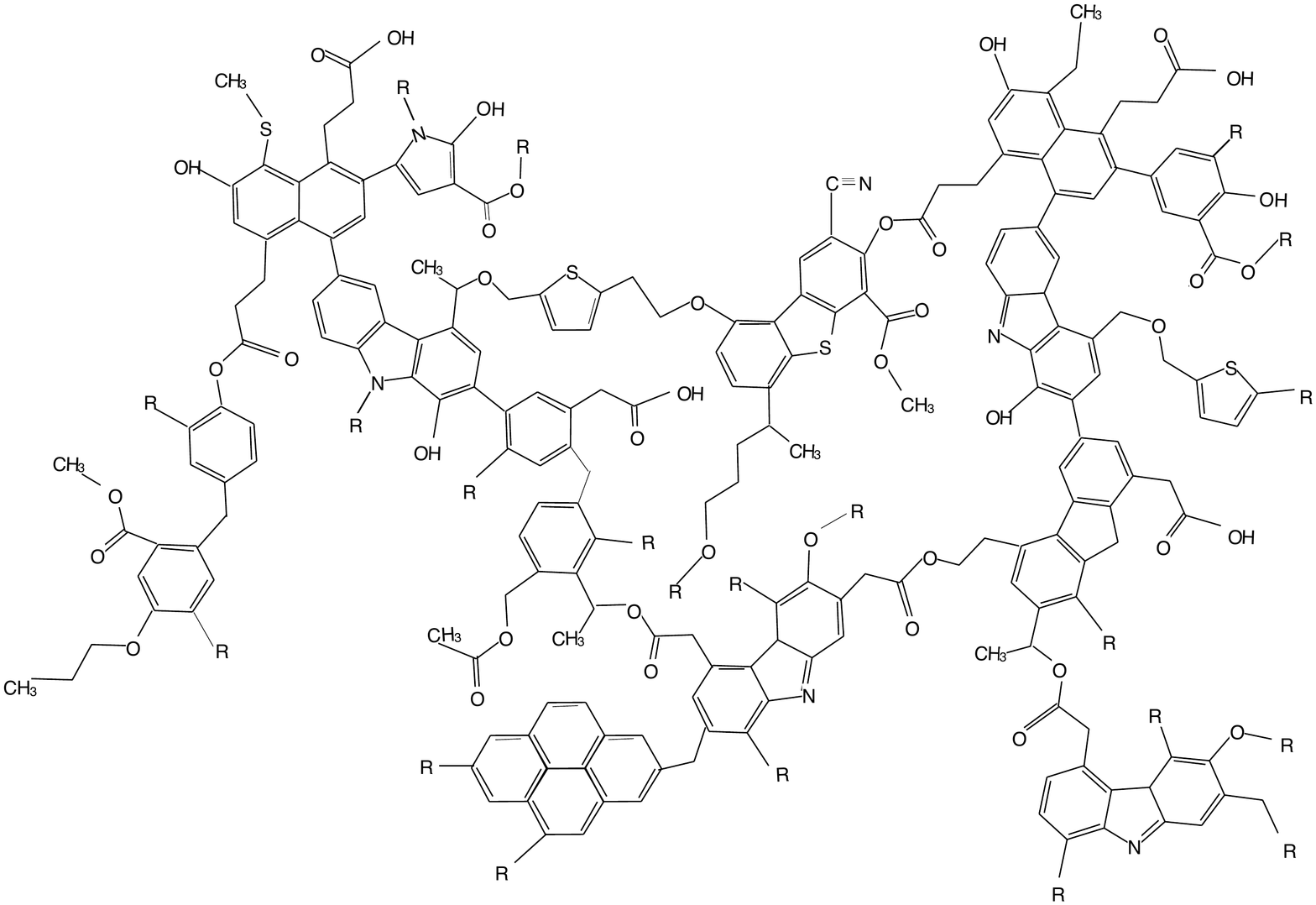}
 \caption{\small Statistical model of the molecular structure of the
   IOM isolated from the Murchison CC. The ``R'' marks the presence of
   an unknown organic groups, called moiety. Adapted from
   \cite{2010M&PS...45.1461D}.}
\label{fig:iom-fig2}
\end{figure}

IOM represents the most abundant (70--90\%) form of carbon isolated
from CCs \citep{1966PNAS...56.1383B}. It is mostly composed by carbon
rings and chains, as shown in Fig.  \ref{fig:iom-fig2}, and contains
several other atoms: O, N, S, and P
\citep{2005LPI....36.1350R,2005GeCoA..69.1085C,2006M&PSA..41.5259R,2010M&PS...45.1461D}. Note,
however, that the molecular structure sketched in the figure should
not be regarded as an organic formula of the IOM, but rather as a
statistical representation of the relative abundances of the organic
bonds in the IOM. In addition, a study by \cite{2011PNAS..10819171C}
showed a molecular relationship between chondritic and cometary
organic solids.  Remarkably, the possible molecular structure of
interstellar refractory organic, namely Poly Aromatic Hydrocarbons
(PAHs), seems to be quite different from that of the IOM
\citep{2004Natur.430..985K,2005LPI....36.1350R}. What is the origin of
this difference is still an open issue: it may hint to a different
origin of the two or to a substantial restructuration of PAHs on the
protosolar grains.

Studies of the molecular deuteration in the IOM have been carried out
by several groups and with a variety of techniques used in organic
chemistry
\citep{1982GeCoA..46...81R,1983GeCoA..47.2199Y,2011IAUS..280..288A}. Also
in this case, high values have been measured, as summarised in
Fig. \ref{fig:iom-fig1}. However, contrarily to the CCs clay minerals
and IDPs, the D/H distribution is not peaked around the terrestrial
value but rather spread from 1.5 to $\sim$3.7$\times10^{-4}$. Besides,
the measured values are larger than those in CCs bulk (dominated by
clay minerals, see above), IDPs and Earth and more similar to what
measured in cometary water (Fig. \ref{fig:iom-fig1}). As discussed in
\S \ref{sec:prot-phase}, the difference between water and organic
deuterium fractionation in protostellar sources, where water is
systematically less deuterated, has been interpreted as due to a
different history of the water and the organic species formation on
the grain surfaces. This may also be the case for the water and
organic material in CCs and IDPs as we will discuss in the following.

Remarkably,
\cite{2006E&PSL.243...15R,2009ApJ...698.2087R,2012GeCoA..96..319R}
have discovered a relation between the D/H ratio measured in the IOM
and the energy of the different C-H bonds, namely aromatic (protonated
carbon in aromatic moieties), aliphatic (in chains linking aromatic
moieties) and benzilic bonds (in the alpha position of the cycles).
In addition to these three bonds, the D/H ratio of the benzilic
carbons attached to aromatic radicals has been measured through
Electron Paramagnetic Resonance (EPR). These radicals are considerably
enriched in deuterium, with D/H = (1.5$\pm$0.5) $\times 10^{-2}$
\citep[cf. Fig. 5; ][]{2008GeCoA..72.1914G}.  This is the largest
deuterium enrichment measured in the primitive objects of the Solar
System.
The occurrence of the so-called "deuterium hot-spots" observed with
the NanoSims technique and that are dispersed in the IOM, is ascribed
to a local micrometer size concentration of these radicals among the
molecular network of the IOM \citep{2009ApJ...698.2087R}.

\begin{figure}[bt]
 \includegraphics[width=8cm]{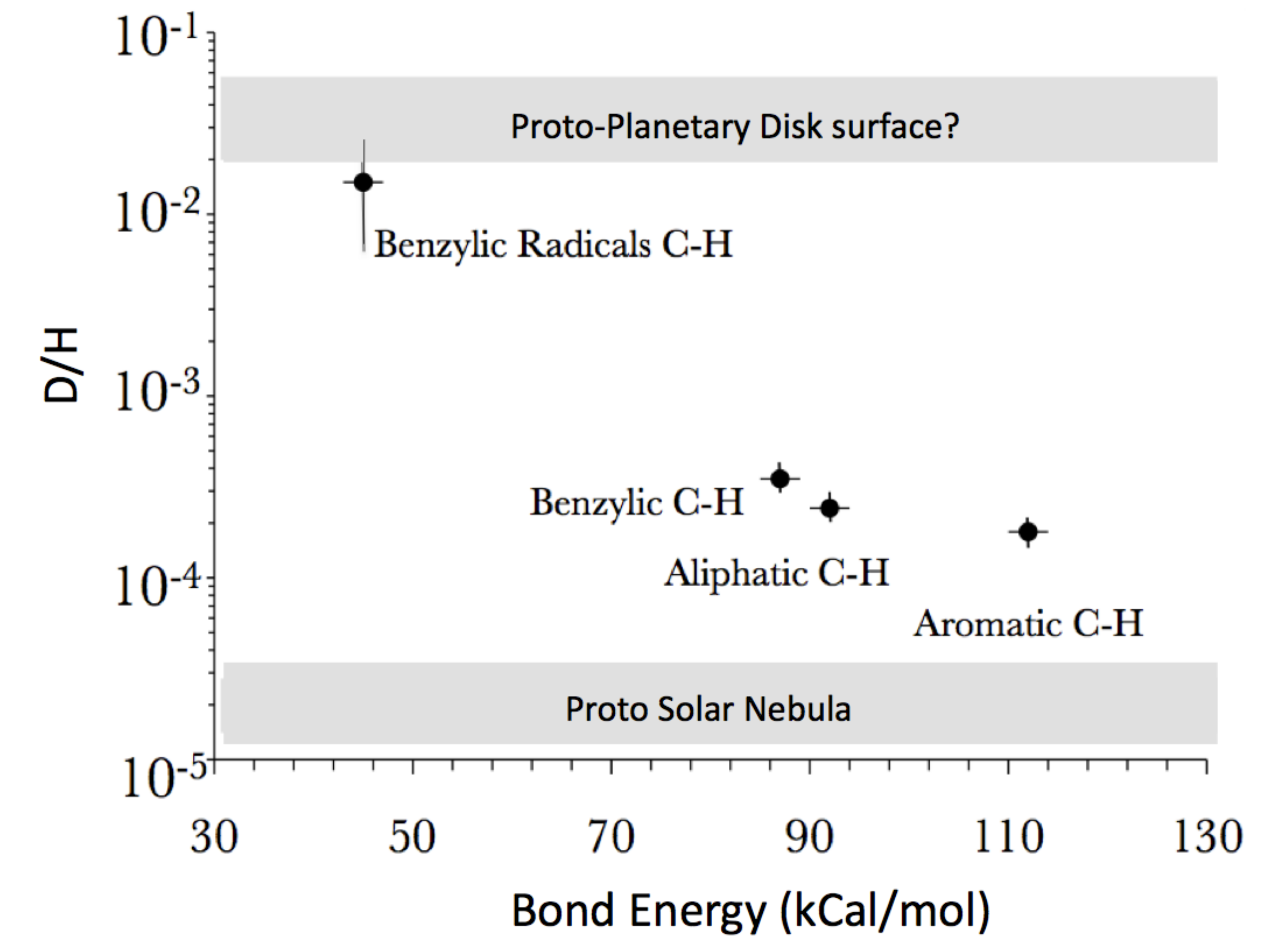}
 \caption{\small Measured D/H ratio in functional groups as function
   of the bond energy of the group.}
\label{fig:iom-fig3} \end{figure}

The relation between the measured D/H and the energy of the bond is
shown in Fig. \ref{fig:iom-fig3}: the larger the bond energy, the
smaller the measured D/H. 
This relation between the energy bond and the deuterium fractionation
led \cite{2006E&PSL.243...15R,2008GeCoA..72.1914G,2010M&PS...45.1461D}
to propose that the deuteration of the IOM occurred during the
protoplanetary disk phase of the Solar System rather than during the
pre-/proto- stellar phase or during the parent body metamorphism
(see \S \ref{sec:solar-nebula}).

\subsection{\textbf{Soluble Organic Compounds}}
\begin{figure}[tb]
 \includegraphics[width=8.5cm]{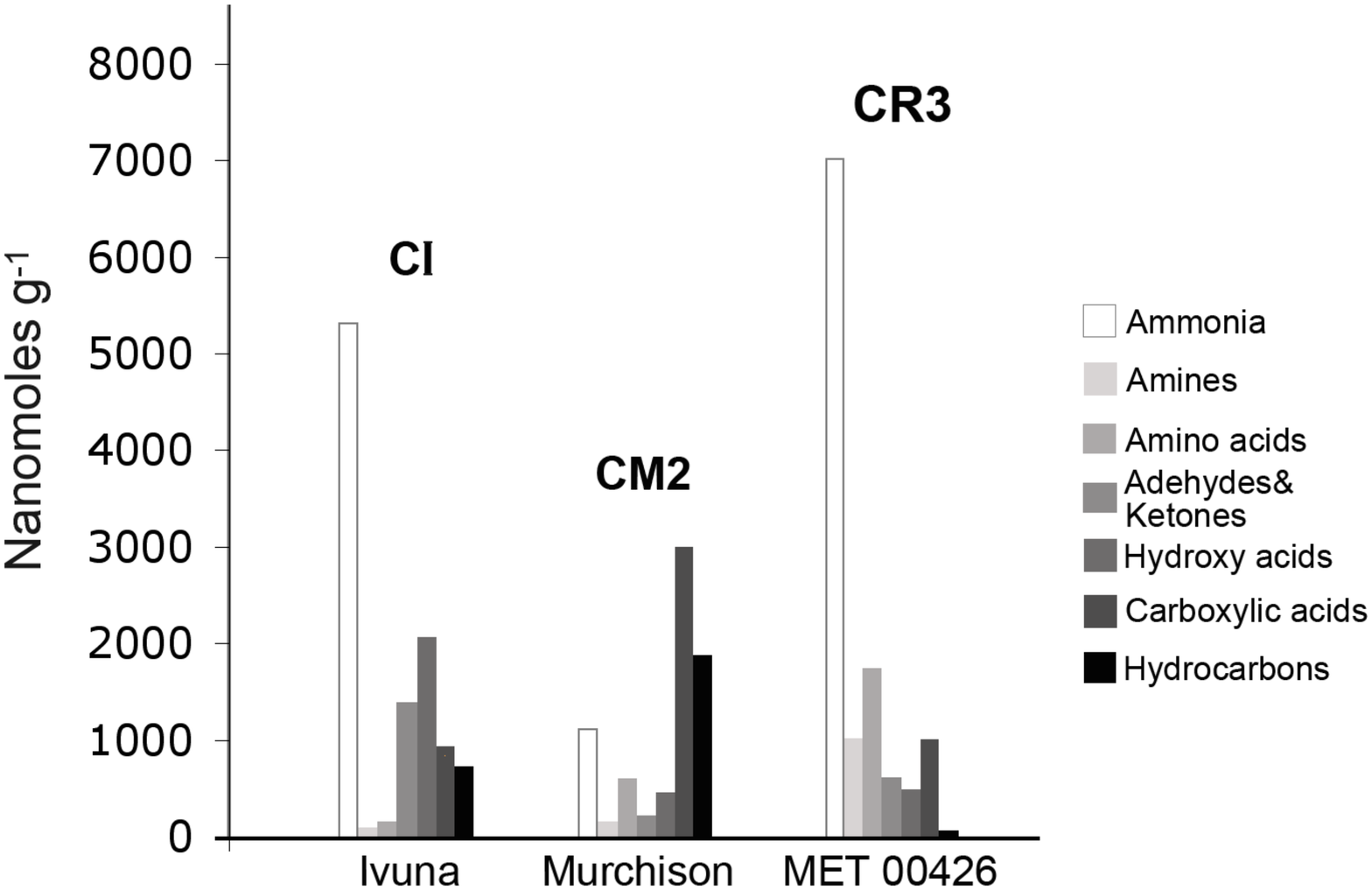}
 \caption{\small Abundance distribution of soluble organic compounds
   in different types of CCs (Carbonaceous Chondrites) meteorites (CI:
   Ivuna-type; CM: Mighei-type; CR: Renazzo-type; MET: Meteorite
   Hills). In general, numbers 1-3 indicate asteroidal aqueous
   alteration extent, where 3 means least altered.}
\label{fig:som-fig1}
\end{figure}

The distribution of the most abundant species of CC soluble compounds
is shown in Fig. \ref{fig:som-fig1}. Besides the ones shown, more
molecules of diverse composition were identified, such as di- and
poly- carboxylic acids, polyhols, imides and amides \citep[][and
references therein]{2006mess.book..625P}, nucleobases
\citep{2011PNAS..10813995C}, and innumerous others are implied by
spectroscopy \citep{2010PNAS..107.2763S}. Their distributions vary
significantly between CCs, e.g., ammonia and amino acids are the most
abundant compounds in Renazzo-type (CR) but hydrocarbons are in
Mighei-type (CM) meteorites.

\begin{figure*}[t]
 \includegraphics[width=20 cm]{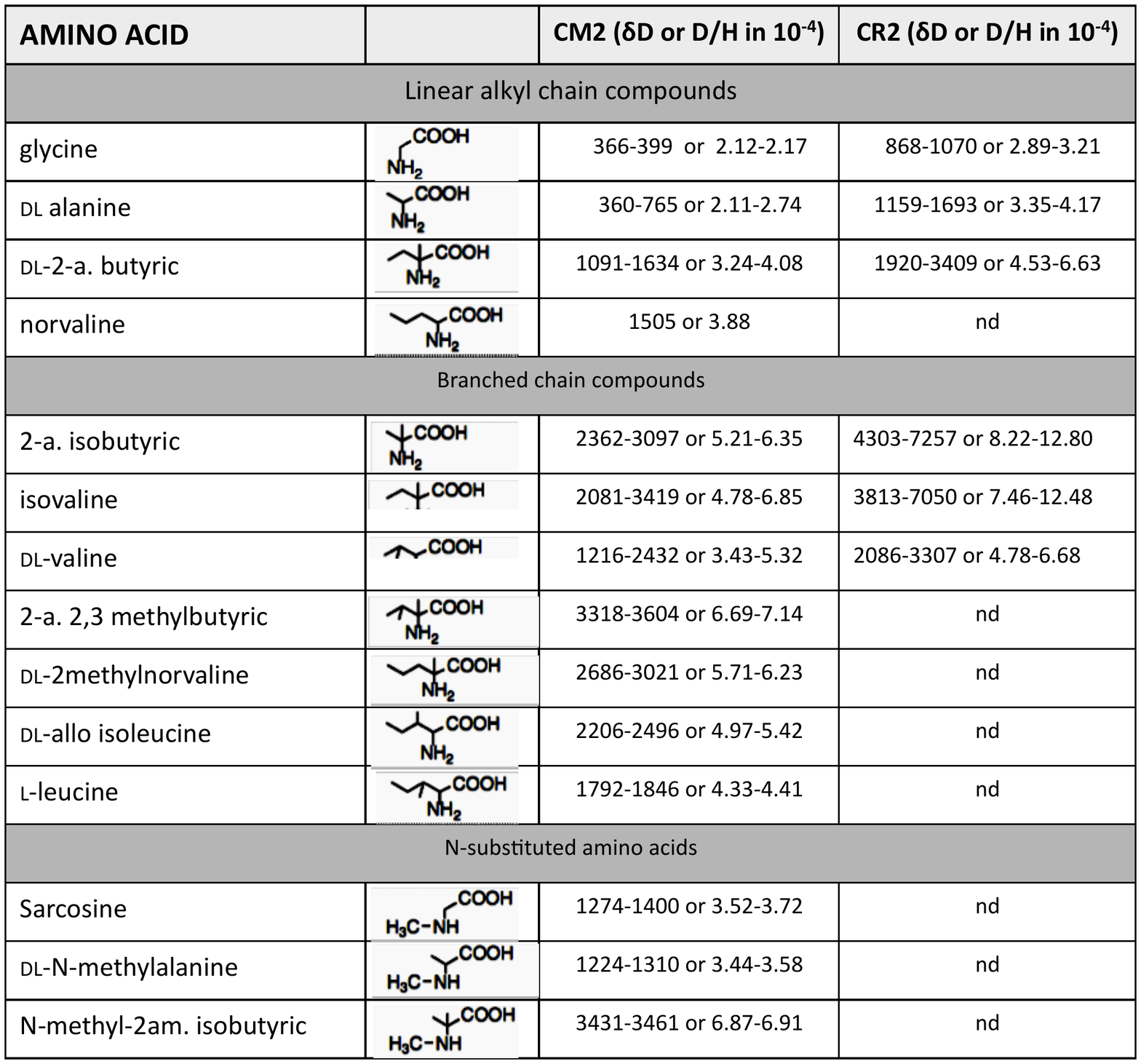}
 \caption{\small Range of $\delta$D~ \textperthousand ~values (see
   Tab. \ref{tab:definitions} for definition) and in D/H units of
   $10^{-4}$ determined for meteoritic 2-amino, and 2-methylamino
   acids, by compound-specific isotopic analyses for CM2 (Murchison,
   Murray and Lonewolf Nunataks 94101) and CR2 (for Graves Nunataks
   95229 and Elephant Morains 92042). Data are from
   \cite{2005GeCoA..69..599P} and \cite{2012M&PS..tmp..201E}. nd= not
   determined.}
\label{fig:som-fig2}
\end{figure*}

Amid this complex suite, amino acids ($aa$) appear to carry a unique
prebiotic significance because of their essential role as constituent
of proteins in the extant biosphere. 
Moreover, meteoritic amino acids may display chiral asymmetry in
similarity with their terrestrial counterparts
\citep[][and references therein]{2011GeCoA..75..645P}, suggesting that abiotic cosmochemical
processes might have provided the early Earth with a primed inventory
of exogenous molecules carrying an evolutionary advantage.
The amino acids of CCs comprise a numerous and diverse group, having linear or
branched alkyl chains and primary as well as secondary amino groups
(Fig. \ref{fig:som-fig2}) in different positions along those
chains. These subgroups may require different pathways of formation
and/or precursor molecules; all are enriched in $^{13}$C, $^{15}$N and
D and appear to have their own isotopic signatures.  This isotopic
diversity was first made evident by the varying $\delta^{13}$C values
detected for Murchison $aa$ (+4.9 to +52.8$\textperthousand$) and the
significant differences found within and between $aa$ subgroups; e.g., the
$^{13}$C content of 2-amino acids (Fig. \ref{fig:som-fig2}) declines
with chain length, as seen for carboxylic acids and alkanes, while 2-,
3-, and 4-amino-, or dicarboxylic amino acids do not have similar
trends. Also 2-methyl-2-amino acids are more enriched in $^{13}$C than
the corresponding 2-H homologues \citep{2004GeCoA..68.4963P}.
The $\delta$D values, determined for a larger number of $aa$ in several
meteorites (Fig. \ref{fig:som-fig2}), show that $aa$ subgroups are
isotopically distinct in regard to this element as well and reveal
additional differences amongst types of meteorites: again, branched $aa$
display the largest $\delta$D values. A higher D-content of branched
alkyl chains was seen also in all other $aa$ sub-groups analyzed, i.e.,
in the dicarboxylic $aa$ and 3-, 4-, 5-amino compounds. 

The locales, precursor molecules and synthetic processes that might
account for meteoritic $aa$ molecular, isotopic and chiral properties
are still being debated and have been only in part elucidated. As for
their synthetic pathways, the first analytical indication came from
the finding in meteorites of similar suites of hydroxy-, and amino
acids, which led \cite{1978Natur.272..443P} to propose a Strecker-type
synthesis for both compound groups:
\begin{figure}[h]
 \includegraphics[width=8cm]{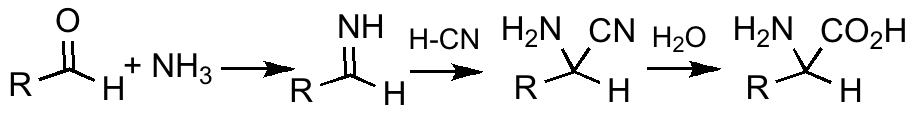}
\label{fig:som-fig4}
\end{figure}

The reaction proceeds via the addition of HCN to aldehydes and ketones
in the presence of ammonia and water; the larger the ammonia
abundance, the larger the expected ratio of NH$_2$-/OH-acids, the
latter being the only compounds formed in ammonia absence. The scheme
agrees with many data from CCs analyses but was further corroborated
recently by enantiomeric excesses observed in the 6C $aa$ isoleucine
of several meteorites \citep{2012PNAS..10911949P}, which demonstrated
the possible origin and survival of enantiomeric excesses from their
precursor aldehyde.

A Strecker synthesis readily accounts for meteoritic $aa$ large D/H
ratios: all reactants necessary for its first phase have been detected
in interstellar clouds and are highly deuterated. The incorporation of
little-altered interstellar species into volatile-rich CC parent
bodies would then provide the subsequent aqueous phase. 

The possible reliance upon minerals by organic materials through their
Solar and pre-Solar formative history is not known and remains one
unexplained feature of chemical evolution. For carbonaceous
meteorites, evidence of CC exposure to liquid water indicates that
aqueous processes and partaking minerals contributed to their final
organic inventory.

\section{\textbf{THE PROTO SOLAR NEBULA MODELS}}\label{sec:solar-nebula}

In this section we discuss separately the models that have been
proposed to explain the measured deuterium fractionation of water and
organics in comets, CCs and IDPs. As in the relevant literature, we
will use the enrichment factor $f$, defined as the ratio between the
water D/H and the H$_2$ D/H of the Proto Solar Nebula (PSN) 
(Tab. \ref{tab:definitions}), which for the terrestrial evaporated
water (VSMOW) is 7.

\subsection{\textbf{Deuterium fractionation of water}}
\begin{figure*}[bt]
\begin{center}
\includegraphics[angle=0,width=12cm]{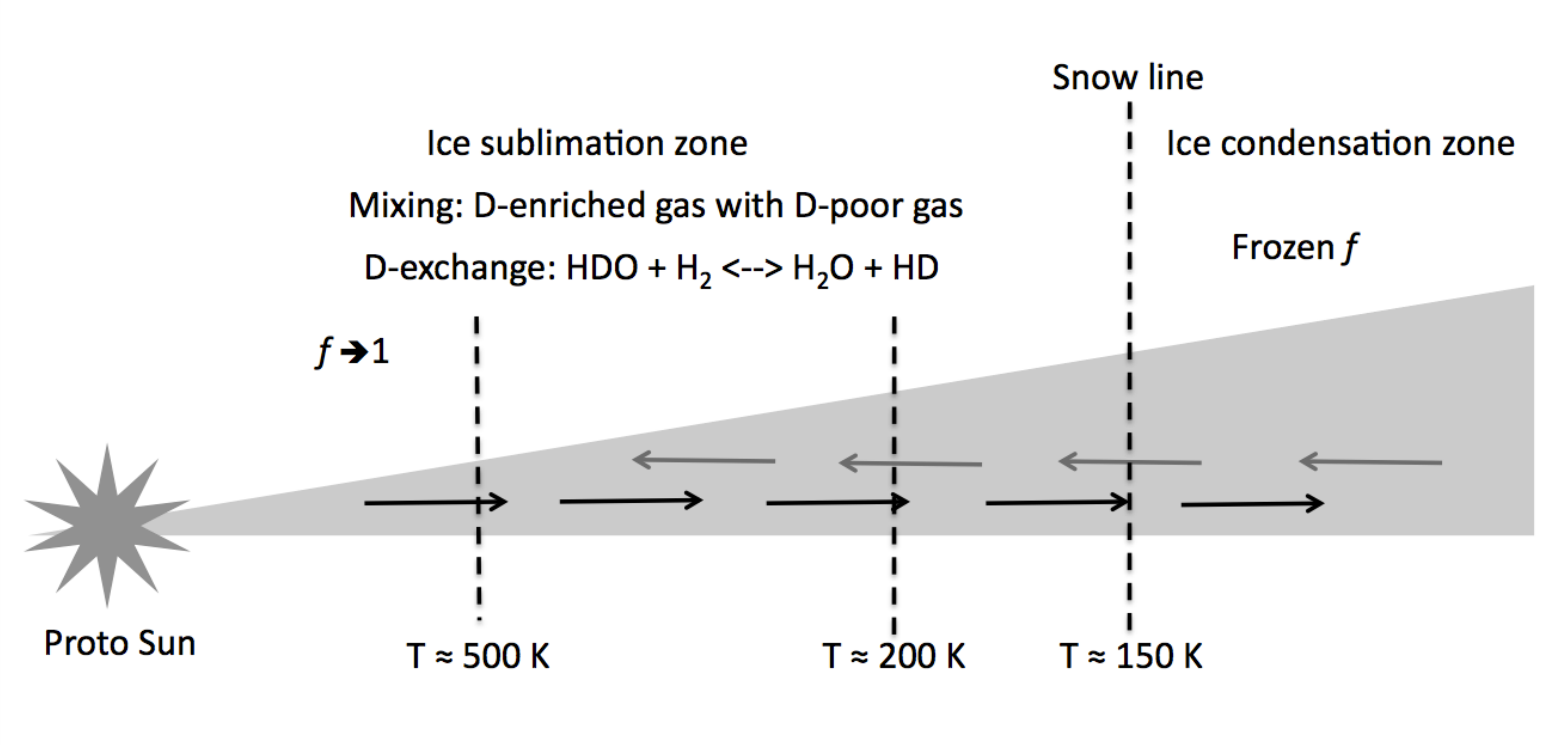}
\caption{\small Cartoon of PSN models. The PSN is described as a
  turbulent disk where matter diffuses inwards and outwards. Inside
  the snow line, wich moves outwards with increasing time, ices
  sublimate injecting D-enriched water vapour in the gas. The water
  vapour then exchanges the D-atoms with H$_2$ via the reaction H$_2$O
  + HD $\rightleftharpoons$ HDO + H$_2$ where $T\ga200$ K. At $T\ga
  500$ K the exchange is very efficient and $f$ tends to unity. When
  the outgoing gas reaches the condensation zone, it freezes out onto
  the grain ices. The acquired D-enrichemnt ($f$) is then conserved
  and inherited by the objects formed at that point.}
\label{SSmodels-fig1}      
\end{center}
\end{figure*}

The PSN, namely the protoplanetary disk out of which the Earth formed,
was initially composed of dust and gas (\S \ref{sec:the-prot-disk})
with a D/H ratio close to the cosmic one ($\sim$ 2.1 $\times$10$^{-5}$
versus $\sim$ 1.6 $\times$10$^{-5}$; Tab. \ref{tab:definitions}).  As
described in \S \ref{sec:set-stage} to \ref{sec:the-prot-disk},
several reactions in the pre-stellar to protoplanetary disk phases
lead to the fractionation of deuterium of H-bearing species, and, in
particular, of the water ice (\S \ref{sec:fract-water}). The
planetesimals that eventually formed comets, the parent bodies of IDPs
and meteorites, and possibly Earth (see the discussion in \S
\ref{sec:terr-water}) were, therefore, covered by icy mantles enriched
in deuterium.  In the warm ($\ga$100--120 K) regions of the PSN,
inside the snow-line, these ices sublimated.  In general, PSN models
aiming to model the fate of the ice
\citep{1999Icar..140..129D,2000Icar..148..513M,2004A&A...414.1165M,2001ApJ...554..391H,2007EM&P..100...43H,2008P&SS...56.1585H,2011ApJ...734L..30K}
assumes that the PSN is a ``diffusional'' disk, in the sense that the
turbulence is entirely responsible for its angular momentum
transport. The first models used the simplest description for the disk
provided by the famous $\alpha$-viscosity 1D (vertical averaged) disk
(Shakura \& Sunyaev 1973). Later versions took into account the 2D
structure of the disk \citep[][]{2001ApJ...554..391H}. The viscosity,
in the numerical models set by the value of $\alpha$, is not only
responsible for the mass accretion from the disk to the star (and,
consequently, the disk thermal structure), but also for the dispersion
of (part of) the angular momentum by diffusing matter outwards.
From a chemical point of view, these models consider only the exchange
of D-atoms between water and molecular hydrogen through the reaction
\citep{1981A&A....93..189G}:
\begin{equation}
\mathrm{H_2O + HD} \rightleftharpoons \mathrm{HDO + H_2}
\label{HDO}
\end{equation}
and consider the isotopic exchange proportional to the reaction rate,
measured by \cite{1994GeCoA..58.2927L}, and the isotopic
fractionation at equilibrium, measured by \cite{1977AREPS...5...65R}.
Therefore, in these models, the deuterium fractionation of water
across the PSN depends on the deuterium exchange in the gas (where 
$T\ga 200$ K) and the mixing of the outwards diffused D-unenriched water
with the inwards diffused D-enriched water.
At high temperatures ($\ga 500$ K) this exchange leads to $f$=1,
namely no water deuterium enrichment. At lower temperatures, the
exchange leads to $f \le 3$, but, being a neutral-neutral
reaction, at temperatures $\la 200$ K the reaction is too slow to play
any role \citep[][]{1999Icar..140..129D}.
Therefore, the deuterium fractionation of water across the PSN, in
these models, depends on the deuterium exchange in the gas (where
$200\la$ T $\la 500$ K) and the mixing of the outwards diffused
D-enriched water (sublimated from the ices) with the D-unenriched
gas. When the gas reaches a temperature equal to that of water
condensation (which is a function of the heliocentric distance and
time), the evolution of $f$ in that point stops. It is then assumed
that the ice D-enrichment is $f$ and that any object formed at that
distance will inherit the same enrichment. This implies that transport
mechanisms of grains, such as turbulent diffusion or gas drag, are
neglected and that there is no evolution afterwards \citep[e.g., in
cometary nuclei during their orbital evolution;][]{Bro12}.  Figure
\ref{SSmodels-fig1} shows a cartoon of the situation.

In summary, these models, based on a strong assumption about the
structure and dynamics of the PSN, depend on two basic parameters: the
$\alpha$ value (or the turbulent viscosity prescription), which
governs the thermal evolution of the disk and the diffusion, and the
initial $f$ profile before the disk evolution starts.
\begin{figure}[bt]
\begin{center}
\includegraphics[angle=-90,width=8.5cm]{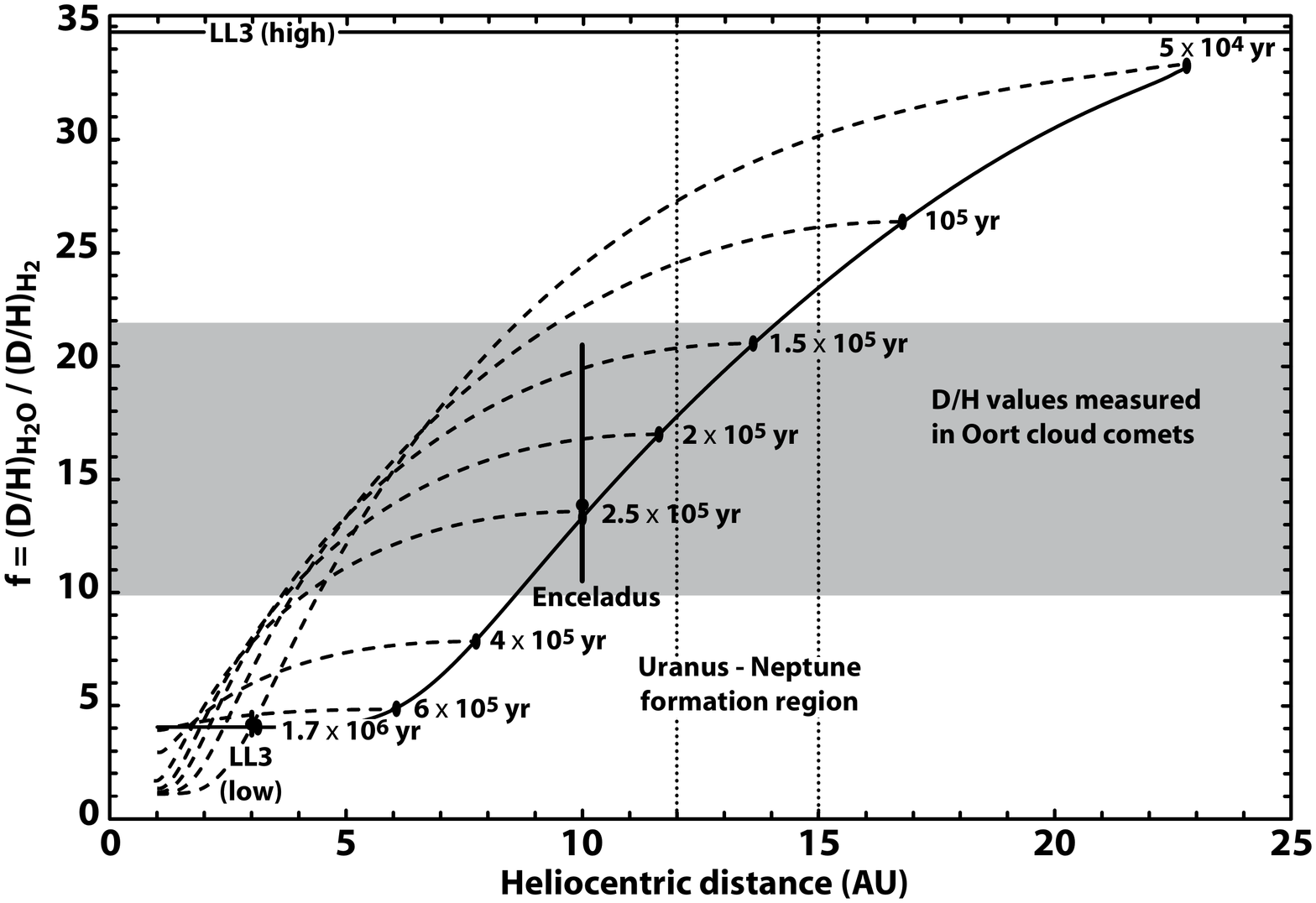}
\caption{\small Enrichment factor $f$ (Tab. 1) as a function of the
  heliocentric distance, as predicted by the \cite{2011ApJ...734L..30K}
  model. $f$ increases in the gas (dashed curves) during the
  PSN diffusive phase evolution (Fig. 14; see text). The $f$ increase
  stops when gas condenses (solid curve), which occurs at different
  times (marked as dots) for different heliocentric distances. The two
  horizontal lines marked with LL3 represent the D/H enrichments in
  LL3 meteorites, low and highly enriched components respectively.
  The vertical line marked Enceladus gives the range of $f$ measured
  in this moon. The vertical dotted lines mark the region between
  Uranus and Neptune, where presumably the Oort cloud comets
  formed. The grey area corresponds to the $f$ values measured in the
  Oort-cloud comets (Sec. \ref{sec:the-comets}).}
\label{SSmodels-fig2}      
\end{center}
\end{figure}
Figure \ref{SSmodels-fig2} shows the evolution of $f$ as a function of
the heliocentric distance as computed by
\cite{2011ApJ...734L..30K}. In this model, the initial $f$ is constant
across the disk and equal to the value found in LL3 meteorites
\citep[D/H=0.9--7.3$\times 10^{-4}$;][]{1998GeCoA..62.3367D}. The
deuterium enrichment profile matches quite well the HDO/H$_2$O values
measured in the Oort-cloud comets (\S \ref{sec:the-comets}), if they
are formed in the Uranus/Neptune region as predicted by the Nice model
\citep{2008Icar..196..258L,Brasser2013}. The computed deuterium
enrichment profile also matches the D/H value measured by the Cassini
spacecraft in the plumes of Saturn's satellite Enceladus at 10 AU,
implying that the building blocks of this satellite may have formed at
the same location as the Oort-cloud comets in the outer PSN
\citep{2009Natur.460..487W,2011ApJ...734L..30K}. However, the same
model cannot reproduce the more recent HDO/H$_2$O value observed
towards the Jupiter-Family comet 103P/Hartley (\S \ref{sec:the-comets}).

\begin{figure}[tb]
\begin{center}
\includegraphics[angle=0,width=8.5cm]{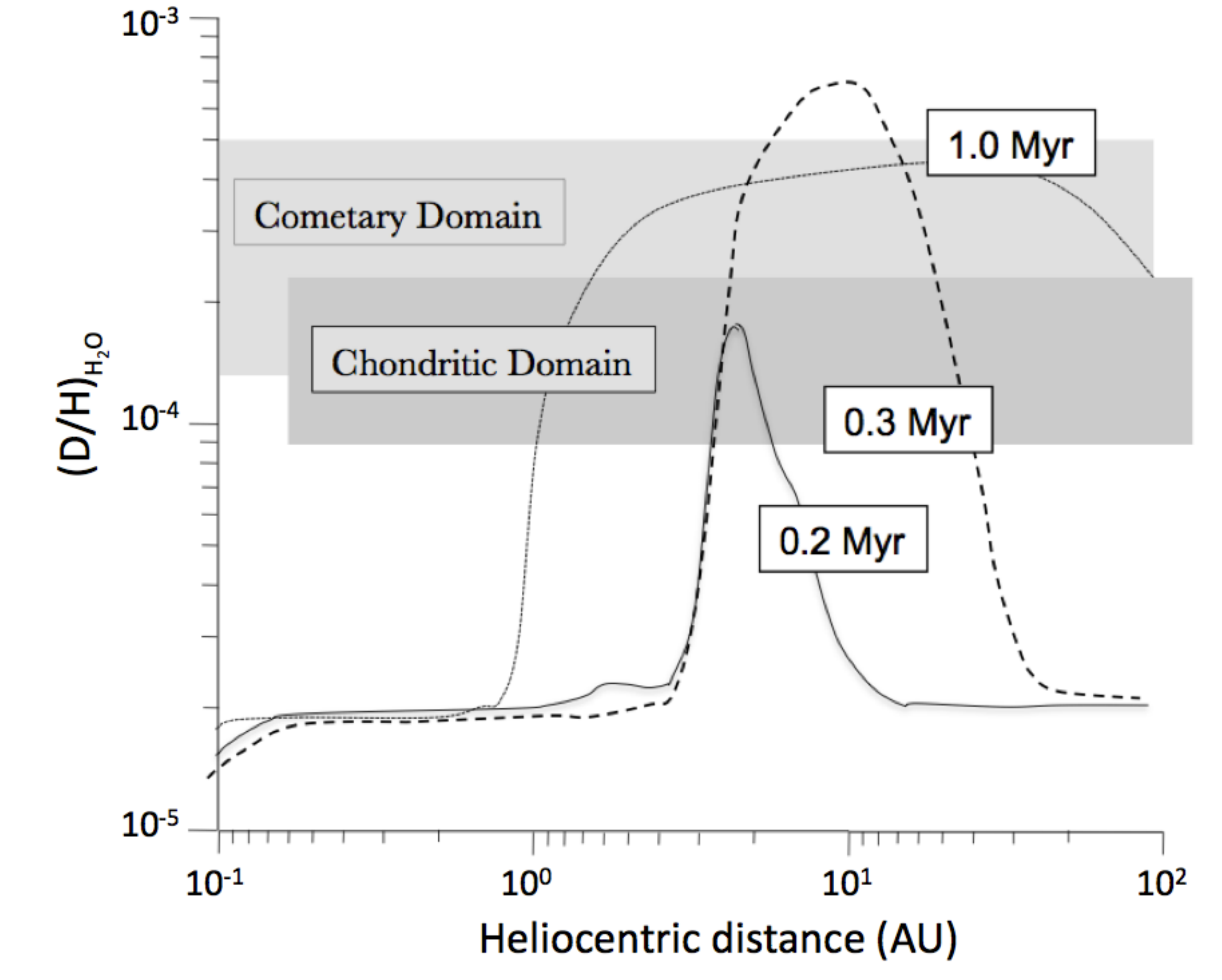}
\caption{\small The radial distribution of D/H in the PSN at
  different epochs of its evolution \citep[adapted from][]{2013Icar..290W}. Jupiter-family comets, presumably formed in the
  trans-Neptunian region, are predicted to have a lower D/H ratio than
  the Oort cloud comets, presumably in majority formed in the
  Uranus/Neptune region.}
\label{SSmodels-fig3}      
\end{center}
\end{figure}
To solve this problem, \cite{2013Icar..290W} developed a new model
whose major differences with the models mentioned above are: (i) the
PSN disk is not an isolated system but it is fed by the infalling
surrounding envelope; (ii) the chemistry is not only governed by the
H$_2$O--H$_2$ isotopic exchange, and other important reactions are
taken into account, following the work by \cite{2010MNRAS.407..232T}
(see also \S \ref{sec:the-prot-disk}). The inclusion of fresh material
from the envelope outer regions leads to a non-monotonic $f$ profile
(Fig. \ref{SSmodels-fig3}), contrarily to what predicted by the
previous studies. This result stands also if the ``old'' chemical
network, namely the isotopic exchange between water and hydrogen, is
adopted.  This non-monotonic gradient in deuteration coud then explain
the HDO/H$_2$O ratio measured towards 103P/Hartley 2 and
45P/Honda-Mrkos-Pajdu\u{s}{\'a}kov{\'a}.

Another important issue, rised by the asymmetric distribution of the
water D/H in CCs and IDPs clays, has been recently addressed by
\cite{2013Icar..223..722J}. In this model, the initial D/H value in
the water ice is assumed to be $3\times$ 10$^{-4}$, the average
observed value in Oort cloud comets (\S \ref{sec:the-comets}). As in
the previous models, iced grains drift inwards and vaporize at the
snow line. Inside the snow line, the water vapor exchanges its D with
H$_2$ and, as a result, the D/H ratio in water vapor
decreases. Reaction kinetics compel isotopic exchange between water
and hydrogen gas to stop at T$\sim$500 K, well inside the snow line
($\sim$200 K). As a consequence, an isotopic gradient is established in
the vapor phase between 200 K and 500 K by the mixing between the
cometary source and the D/H ratio established at 500 K. The maximum
value reached by the D/H ratio around the snow line is 1.2
$\times$10$^{-4}$. A fraction of this water having a D/H ratio of 1.2
$\times$10$^{-4}$ can be transported beyond the snow line via
turbulent diffusion where it (re)condenses as ice grains that are then
mixed with isotopically comet-like water (D/H $\sim$ 3 $\times$
10$^{-4}$). Thus, the competition between outward diffusion and net
inward advection establishes an isotopic gradient, which is at the
origin of the large isotopic variations in the carbonaceous chondrites
and their asymmetric distribution. Under this scenario, the calculated
probability distribution function of the D/H ratio of ice mimics the
observed distribution reported for CCs in Fig. \ref{fig:iom-fig1}. In
this interpretation, the minimum D/H ratio of the distribution
($\sim$1.2 $\times$ 10$^{-4}$) corresponds to the composition of water
(re)condensed at the snow line. The distribution function depends on
several parameters of the model that are discussed by
\cite{2013Icar..223..722J}.

Interestingly, the presence of water on Earth, which exhibits a bulk
(i.e., including surface water and mantle hydroxyls) D/H value of 1.49
$\pm$0.03 $\times$ 10$^{-4}$\citep{Lecuyer1998}, can simply be
accounted for by formation models advocating that terrestrial planets
accreted from embryos sharing a CC composition
\citep{2000M&PS...35.1309M}. In this case, the contribution of comets
to the Earth's water budget would be relatively negligible.

\subsection{\textbf{Deuterium fractionation of organics}}\label{sec:fract-organ}
As discussed in \S \ref{sec:meteorites}, organic matter shows a
systematic larger deuterium fractionation with respect to water
(Fig. \ref{fig:iom-fig1}). A precious hint of what could be the reason
for that is given by the relation between the energy of the functional
group bond and the measured D/H value (Fig. \ref{fig:iom-fig3}). To
explain this behaviour, several authors
\citep{2006E&PSL.243...15R,2008GeCoA..72.1914G,2010M&PS...45.1461D}
suggested that D-atoms are transmitted from HD to the organics via the
H$_2$D$^+$ ions present in the disk \citep[][see also \S
\ref{sec:the-prot-disk}]{2005A&A...440..583C,2008ApJ...681.1396Q,Willacy_Woods09},
basically following the scheme of
Fig. \ref{fig:D-chemistry-scheme}. Laboratory experiments add support
to this hypothesis. \cite{2011GeCoA..75.7522R} experimentally measured the
exchange rate of D-atoms from H$_2$D$^+$ to organic molecules
containing benzylic, aliphatic and aromatic bonds, finding relative
ratios as measured in the IOM. Extrapolating to the typical conditions
of protoplanetary disks, namely scaling to the theoretical ion
  molecular ratio (i.e. to the H$_2$D$^+$/H$_2$ ratio) at the surface
  of the disk, these authors estimated that the exchange is very fast
($10^{4\pm1}$yr) and, therefore, a plausible mechanism to occur.
Therefore, the fact that the D/H distribution for the IOM is
systematically enriched in D compared with that for water
(cf. Fig. \ref{fig:iom-fig1}) could result from reactions in the gas
phase and not connected with a fluid circulating in parent body
meteorites during a hydrothermal episode. Indeed, all known isotopic
reactions of organic molecules with a fluid should have resulted in
the opposite distribution i.e., a systematic depletion in the organic
D/H ratio compared with that of clays.

\section{\textbf{FOLLOWING THE UNROLLED DEUTERIUM THREAD}}\label{sec:summaryD}
\begin{figure*}[tb]
\begin{center}
 \includegraphics[width=12cm, angle=-0]{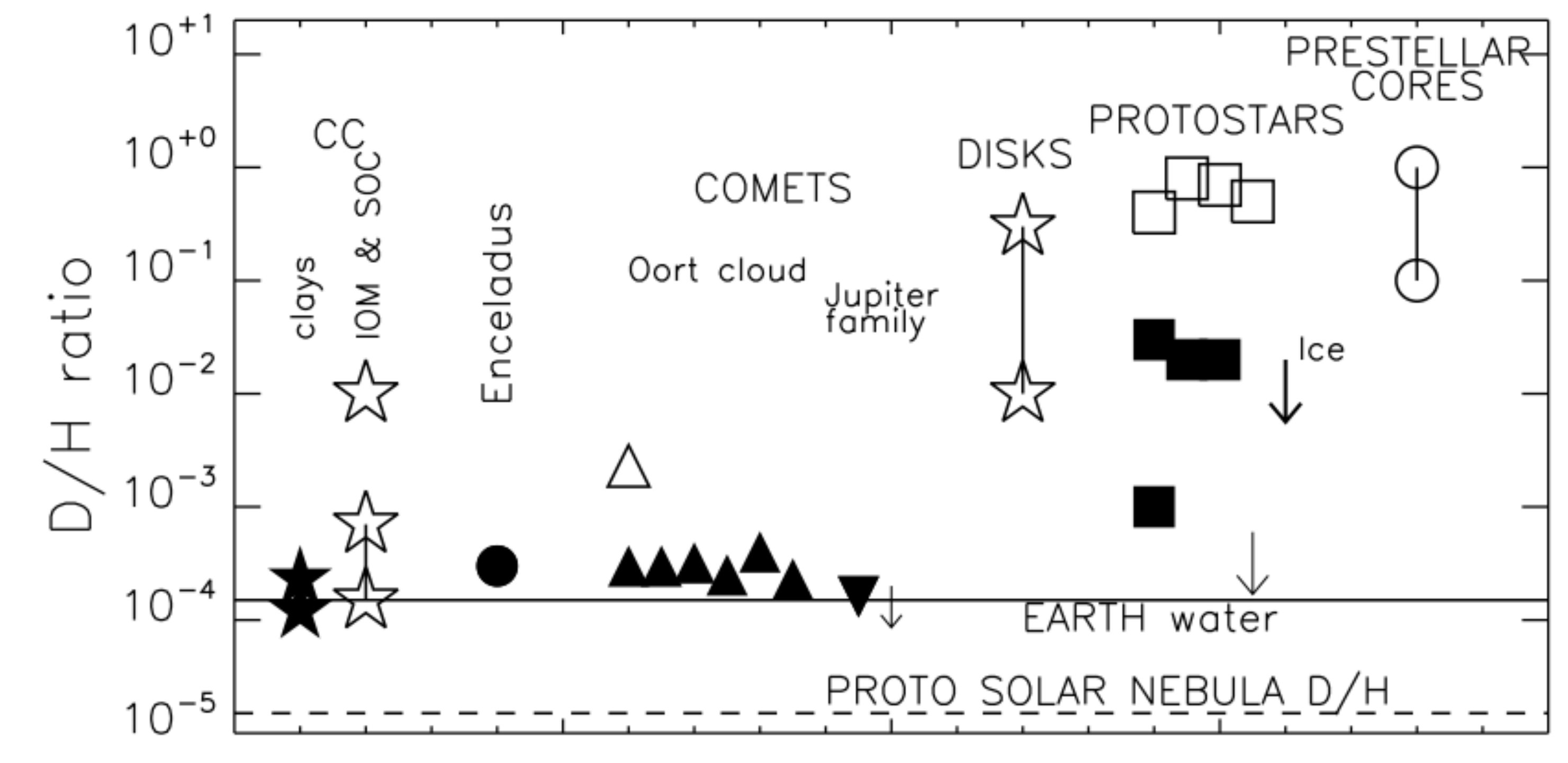}
 \caption{\small D/H ratio in the Solar System objects and
   astronomical sources. The filled symbols refer to water measures
   whereas open symbols refer to organic matter. The different objects
   are marked.}
\label{fig:DH-summary}
\end{center}
\end{figure*}
\subsection{\textbf{Overview of the deuterium fractionation through the different phases}}\label{sec:overview-meas-deut}
Figure \ref{fig:DH-summary} graphically summarises the measured values
of the D/H ratio from pre-stellar cores up to Solar System
bodies, whereas Table \ref{tab:DH-summary} lists the measured D/H
values and the main characteristics of the observations of Solar
System bodies.
\begin{table*}[bth]
  \begin{tabular}[h]{|lll|}
    \hline \hline
    {\bf Object} & {\bf D/H } & {\bf Note} \\
    & [$\times 10^{-4}$] & \\
    \hline
    Oort-Cloud Comets$^a$     &  $\sim 3$ & Measured in seven comets.\\
    Jupiter-Family Comets$^a$ &  $\leq 2 $ & Measured in two
    comets: 1.61$\pm$0.24 and $\leq$2. \\
    \hline
    IDPs$^a$                        &  1.5--3.5 & The D/H distribution
    is peaked at 1.3, with a shoulder from 0.8 to 3.5\\
    & & (it mimics the CC clays
    D/H distribution).\\
    CC clays$^a$                    &  1.2--2.3 & The D/H distribution
    is peaked at 1.4, with a shoulder from 1.7 to 2.3.\\
    CC IOM$^b$                & 1.5--3.5 & The D/H
    distribution does not have a peak.\\
    CC IOM functional groups$^b$ & 1.5--5.5  & The larger the
    functional group bonding energy the larger the D/H.\\
    CC IOM deuterium hot spots$^b$   & $\sim$100 & Micrometer size spots in the IOM
    networks where D/H is highly enhanced.\\
    CC SOC$^b$     &  2.1--7.1 & Branched $aa$ show the largest D/H.\\
    \hline
  \end{tabular}
  \caption{\small  Main characteristics of deuteration fractionation in comets
    and carbonaceous chondrites (CC). For explanations and references, see
    Secs. \ref{sec:the-comets} --
    \ref{sec:meteorites}. Notes: $^a$HDO/H$_2$O;
    $^b$deuterium fractionation in organics; }
  \label{tab:DH-summary}
\end{table*}
Overall, the figure and the table, together with
Fig. \ref{fig:iom-fig1}, raise a number of questions, often
found in the literature too, which we address here.
\\

\noindent {\it 1. Why is the D/H ratio systematically lower in Solar
  System bodies than in pre- and proto- stellar objects? When and how
  did this change occur?}\\
May be here the answer is, after all, simple. All the Solar System
bodies examined in this chapter (meteorites and comets) were
originally at distances less than 20 AU from the Sun. The D/H
measurements of all pre- and protostellar objects examined in this
chapter (pre-stellar cores, protostars and protoplanetary disks) refer
to distances larger than that, where the temperatures are lower and,
consequently, the deuteration is expected to be much larger (\S
\ref{sec:the-chem-proc}). Therefore, this systematic difference may
tell us that the PSN did not undergo a global scale re-mixing of
the material from the outer ($\ga 20$ AU) to the inner regions. There
are exceptions, though, represented by the ``hot spots'' in
meteorites, which, on the contrary, may be the only representatives of
this large scale re-mixing of the material in the PSN.
\\

\noindent {\it 2. Why does organic material have a systematically
  higher D/H
  value than water, regardless the object and evolutionary phase?}\\
The study of the deuterium fractionation in the pre-stellar cores (\S
\ref{sec:the-pre-stell}) and protostars (\S \ref{sec:prot-phase}) has
taught us that the formation of water and organics in different epochs
is likely the reason why they have a different D/H ratio
(Fig. \ref{fig:protostar-1}). Water ices form first, when the density
is relatively low ($\sim 10^4$ cm$^{-3}$) and CO not depleted yet, so
that the H$_2$D$^+$/H$_3^+$ ratio is moderate (\S
\ref{sec:the-chem-proc}). Organics form at a later stage, when the
cloud is denser ($\ga 10^5$ cm$^{-3}$) and CO (and other neutrals)
condenses on the grain surfaces, making possible a large enhancement
of the H$_2$D$^+$/H$_3^+$. Can this also explain, {\it mutatis
  mutandis}, the different deuterium fractionation of water and
organics measured in CCs and comets? After all, if the PSN cooled down
from a hot phase, the condensation of volatiles would follow a similar
sequence: oxygen/water first, when the temperature is higher, and
carbon/organics later, when the temperature is lower. This would lead
to deuterium fractions lower in water than in organics, regardless
whether the synthesis of water and organic was on the gas phase or on
the grain surfaces. Obviously, this is at the moment speculative but a
road to explore, in our opinion. We emphasise that this ``different
epochs formation'' hypothesis is fully compatible with the theory
described in \S \ref{sec:fract-organ} of the origin of the organics
deuteration from H$_2$D$^+$ \citep{2011GeCoA..75.7522R}.
\\

\noindent {\it 3. Why do most comets exhibit higher D/H water values, about a factor 2, than CCs and IDPs?}\\
A possible answer is that comets are formed at distances, $\ga5$ AU,
larger than those where the CCs originate, $\la 4$ AU
(Tab. \ref{tab:definitions}; see also \S 9 ). A little more
puzzling, though, is why the large majority of IDPs have D/H values
lower than the cometary water, although at least 50\% IDPs are
predicted to be cometary fragments (Nesvorny et al. 2010). The new
Herschel observations of JFC indicate indeed lower D/H values than
those found in OCC (Tab. \ref{tab:Dcomets}; \S \ref{sec:the-comets}),
so that the observed IDPs D/H values may be consistent with the
cometary fragments theory. In conclusion, the D/H distribution of CCs and
IDPs is a powerful diagnostic to probe the distribution of their
origin in the PSN.
\\

\noindent {\it 4. Why does the D/H distribution of the IDPs, which are
  thought to be fragments of comets, mimic the CCs bulk D/H
  distribution? And why is it asymmetric, namely with a shoulder
  towards the large D/H values?}\\
As written above, the D/H ratio distribution depends on the
distribution of the original distance from where CCs and IDPs come
from or, in other words, their parental bodies. CCs are likely
fragments of asteroids from the main belt and a good fraction is of
cometary origin. IDPs are fragments from JFC and main belt
asteroids. Their similar D/H distribution strongly suggests that the
mixture of the two classes of objects is roughly similar, which argue
for a difference between the two, more in terms of sizes than in
origin. Besides, the asymmetric distribution testifies that a majority
of CCs and IDPs originate from closer to the Sun parent bodies
\citep[see also][]{2013Icar..223..722J}.
\\

\noindent {\it 5. Why is the D/H ratio distribution of CC organics and
  water so different?}\\
There are in principle two possibilities: water and organics formed in
different locations at the same time and then were mixed together or,
on the contrary, they were formed at the same heliocentric distance
but at (slightly) different epochs. The discussion in point 2 would favor
the latter hypothesis, although this is speculative at the moment.  If
this is true, the D/H distribution potentially provides us with their
history, namely when each of the two components formed. Again, being
CC organics more D-enriched than water, they were formed at later stages.

\subsection{\textbf{Comparison between PPD and PSN
    models: need for a paradigma shift?}}\label{sec:comp-models}
There are several differences between the Proto Planetary disks (PPDs)
and Proto Solar Nebula (PSN) models. A first and major one is that PSN
models assume a dense and hot phase for the solar PPD. For
example, temperatures higher than 1000 K persist for $\sim 10^5$ yr at
3 AU and $\sim 10^6$ yr at 1 AU in the Yang et al. (2013)
model. Generally, models of PPDs around Solar-type stars, on the
contrary, never predict such high temperatures at those distances.
Second, PPD models consider complex chemical networks with some
horizontal and vertical turbulence which modifies, though not
substantially, the chemical composition across the disk. On the
contrary, in PSN models, chemistry networks are generally very
simplified, whereas the turbulence plays a major role in the final
D-enrichment across the disk.  However, since the density adopted in
the PSN models is very high, more complex chemical networks have a
limited impact \citep{2013Icar..290W}.
Third, the above mentioned PSN models do not explicitly compute the
dust particle coagulation and migration (if not as diffusion),
and gas-grains decoupling whereas (some) disk models do (see Turner et
al. and Testi et al., this volume).

In our opinion, the most significant difference is the first one
and the urgent question to answer is: what is the good description of
the PSN? A highly turbulent, hot and diffusive disk or a cooler and
likely calmer disk, as the ones we see around T Tauri stars? Or
something else?

The community of PSN models have reasons to think that the PSN disk
was once hot. A major one is the measurement of depletion of
moderately volatile elements (more volatile than silicon)
in chondritic meteorites (Palme et al. 1988), which can be explained
by dust coagulation during the cooling of an initially hot ($\ga 1400$
K, the sublimation temperature of silicates) disk in the terrestrial
planet formation region \citep[$\la 2$
AU;][]{1974Sci...186..440L,1977Icar...30..447C,1996ApJ...472..789C}. The
explanation, though, assumes that the heating of the dust at $\ga
1400$ K was global in extent, ordered and systematic. Alternatively,
it is also possible that it was highly localised and, in this case,
the hot initial nebula would be not necessary. For example, the
so-called X-wind models assume that the hot processing occurs much
closer to the star and then the matter is deposited outwards by the
early Solar wind
\citep{1996Sci...271.1545S,2000prpl.conf..789S,2001ApJ...548.1051G}.
In addition, the detection of crystalline silicates in comets
\citep{2004ApJ...612L..77W} has been also taken as a prove that the
Solar System passed through a hot phase \citep{2007A&A...466L...9M}.

For sure, T Tau disks do not show the high temperatures ($\ga 1400$ K
at $r\sim 2$ AU) assumed by the PSN model. These temperatures are
predicted in the midplane very close ($\la 0.1$ AU) to the central
star or in the high atmosphere of the disk, but always at distances
lower than fractions of AU \citep{2010MNRAS.407..232T}.  Even the most
recent models of very young and embedded PPDs, with or without
gravitational instabilities \citep[see the review
by][]{2012A&ARv..20...56C}, predict temperatures much lower than
$\sim1400$ K.
One can ask whether the hot phase of the PSN could in fact be the hot
corino phase observed in Class 0 sources (\S
\ref{sec:prot-phase}). Although the presently available facilities do
not allow to probe regions of a few AUs, the very rough extrapolation
of the temperature profile predicted for the envelope of
IRAS16293-2422 (the prototype of Class 0 sources) by
\cite{2010A&A...519A..65C} gives $\sim$1000 K at a few AUs. The
question is, therefore: should we not compare the PSN with the hot
corino models rather than the protoplanetary disks models? At
present, the hot corino models are very much focused only on the $\ga
10$ AU regions, which can be observationally probed, and only consider
the gas composition.  Should we not start, then, considering what
happens in the very innermost regions of the Class 0 sources and study
the dust fate too?  If so, the link with the deuterium
fractionation that we observe in the hot corino phase
may become much more relevant in the construction of realistic PSN
models.  Last but not least, the present PSN models are based on the
transport and mixing of material with different initial D/H water
because of the diffusion, responsible for the angular momentum
dispersion. In Class 0 sources, though, the dispersion of the angular
momentum is thought to be mainly due to the powerful jets and outflows
and not by the diffusion of inward/outward matter. Moreover, during
the Class 0 phase, material from the cold protostellar envelope
continues to rain down onto the central region and the accreting disk
(Z.-Y. Li et al., this volume), replenishing them by highly deuterated
material.The resulting D/H gradient across the PSN may, therefore, be
different than that predicted by current theories.

\subsection{\textbf{Where does the terrestrial water come
    from?}}\label{sec:terr-water}
Several reviews discussing the origin of the terrestrial water are
present in the literature \citep[][van Dishoeck et al, this
volume]{2009P&SS...57.1338H,2012AREPS..40..251M,2012A&ARv..20...56C},
so that we will here just summarise the points emphasising the open
issues.

Let first remind that the water budget on Earth is itself subject of
debate. In fact, while the lithosphere budget is relatively easy to
measure \citep[$\sim 10^{-4}$ M$_\earth$;][]{Lecuyer1998}, the water
contained in the mantle, which contains by far the largest mass of our
planet, is extremely difficult to measure and indirect probes, usually
noble gases, are used for that \citep[e.g.,][]{1983Natur.303..762A},
with associated larger error bars. It is even more difficult to
evaluate the water content of the early Earth, which was likely more
volatile-rich than at present \citep{1996E&PSL.144..577K}. The most
recent estimates give $\sim 2\times 10^{-3}$ M$_\earth$
\citep{2012E&PSL.313...56M}, namely 20 times larger than the value of
the lithosphere water. If the water mantle is the dominant water
reservoir, as it seems to be, then the D/H value of the terrestrial
oceans may be misleading if geochemical processes can alter
it. Evidently, measurements of the mantle water D/H is even
more difficult. The last estimates suggest a value slightly lower than
the terrestrial oceans water. In this chapter, given this uncertainty
on the Earth bulk water content and D/H, we adopted as reference the
evaporated ocean water D/H, the VSWOM (Tab. \ref{tab:definitions}).

The ``problem'' of the origin of the terrestrial oceans rises because,
if Earth was formed by planetesimals at $\sim$1 AU heliocentric
distance, they would have been ``dry'' and no water should exist on
Earth. One theory, called ``late veneer'', assumes that water was
brought after Earth formed by, for example, comets
\citep{1992AdSpR..12....5D,1995Icar..116..215O}. This theory is based
on the assumptions that the D/H cometary water is the same than the
Earth water D/H, but, based on observations towards comets (\S
\ref{sec:the-comets}), this assumption is probably wrong. 
The second theory, based on the work by \cite{2000M&PS...35.1309M},
assumes that a fraction of the planetesimals that built the Earth came
from more distant (2--4 AU) regions and were, therefore,
``wet''. Dynamical simulations of the early Solar System evolution
have add support to this theory
\citep{2000M&PS...35.1309M,2012A&A...546A..18M,2009Icar..203..644R},
challenging at the same time the idea that the flux of late veneer
comets and asteroides could have been large enough to make up the
amount of water on Earth. Moreover, the D/H value measured in CCs (\S
\ref{sec:meteorites}) adds support to this theory. The recent findings
by \citeauthor{2012Sci...337..721A} would argue for a large
contribution of a group of CCs, the CI type.

In summary, the origin of terrestrial water is still a source of intense debate.

\section{\textbf{SUMMARY AND FUTURE PROSPECTS}}\label{sec:conclusions}
In this chapter we have established a link between the various phases
in the process of Solar-type star and planet formation and our Solar
System. This link has been metaphorically called the Ariadne's thread
and it is represented by the deuterium fractionation
process. Deuterium fractionation is active everywhere in time and
space, from pre-stellar cores, to protostellar envelopes and hot
corinos, to protoplanetary disks. Its past activity can be witnessed
by us today in comets, carbonaceous chondrites and interplanetary dust
particles. Trying to understand and connect this process in the
various phases of star and planet formation, while stretching the
thread to our Solar System, has opened new horizons in the quest of
our origins. The ultimate goal is to chemically and physically connect
the various phases and identify the particular route taken by our
Solar System. The steps toward this goal are of course many, but much
will be learned just starting with the following important points:

\begin{enumerate}
\item Bridge the gap between pre-stellar cores and protoplanetary
  disks and understand how different initial conditions affect the
  physical and chemical structure and evolution of protoplanetary
  disks.
\item Study the reprocessing of material during the early stages of
  protoplanetary disks; in particular, can self-gravitating accretion
  disks, thought to be in present in the hot corino/protostellar
  phase, help in understanding some of the observed chemical and
  physical features of more evolved protoplanetary disks and our Solar
  System?
\item Study the reprocessing of material (including dust coagulation
  and chemical evolution of trapped ices) throughout the
  protoplanetary disk evolution; in particular, which conditions do
  favour the production of the organic material observed in pristine
  bodies of our Solar-System?
\item Compare PSN models with protoplanetary disk models and include
  the main physical and chemical processes, in particular dust
  coagulation and ice mantle evolution.
\end{enumerate}

The future is bright thanks to the great instruments available now and
in the near future (e.g., ALMA and NOEMA, the ESA ROSETTA mission) and
the advances in techniques used for the analysis of meteoritic
material. Trying to fill the D/H plot shown in
Fig.\ref{fig:DH-summary} with new observations is of course one
priority, but this needs to proceed hand in hand with developments in
theoretical chemistry and more laboratory work.  For sure, one lesson
has been learned from this interdisciplinary work: our studies of the
Solar System and our studies of star/planet forming regions represent
two treasures which cannot be kept in two different coffers. It is now
time to work together for a full exploitation of such treasures and
eventually understand our astrochemical heritage.
\\

\textbf{ Acknowledgments.} C. Ceccarelli acknowledges the financial
support from the French Agence Nationale pour la Recherche (ANR)
(project FORCOMS, contract ANR-08-BLAN-0225) and the French spatial
agency CNES. P. Caselli acknowledges the financial support of the
European Research Council (ERC; project PALs 320620) and of successive
rolling grants awarded by the UK Science and Technology Funding
Council.  O. Mousis acknowledges support from CNES. S. Pizzarello
acknowledges support through the years from the NASA Exobiology and
Origins of the Solar System Programs. D. Semenov acknowledges support
by the {\it Deutsche Forschungsgemeinschaft} through SPP~1385: ``The
first ten million years of the solar system - a planetary materials
approach'' (SE 1962/1-1 and 1-2).  This research made use of NASA's
Astrophysics Data System.  We wish to thank A. Morbidelli,
C. Alexander and L. Bonal for a critical reading of the manuscript. We
also thank an anonymous referee and the Editor, whose comments helped
to improve the chapter clarity.

\bigskip

\bibliographystyle{ppvi_lim1.bst}

\end{document}